\documentclass[12pt]{article}
\usepackage{amsmath,amssymb}
\usepackage{color}
\usepackage{graphicx}  
\usepackage{appendix}
\def\e{\varepsilon}

\def\8{\infty}

\newtheorem {lemma}{Lemma}
\newtheorem {cor}{Corollary}

\newtheorem {prop}{Proposition}

\usepackage{xcolor}

\makeatletter

\@addtoreset{equation}{section}
\makeatother
\begin{document}

\title{Exact and asymptotic features of the edge density profile for the one component plasma in two dimensions}

\author{T. Can\footnote{
Department of Physics, University of Chicago, 929 57th St, Chicago, IL 60637}
,
P. J. Forrester\footnote{
Department of Mathematics and Statistics, The University of Melbourne, Victoria 3010, Australia}
,
G. T\'ellez\footnote{
Departamento de F\'isica, Universidad de Los Andes, Bogot\'a, Colombia} 
,
P. Wiegmann\footnote{
Department of Physics, University of Chicago, 929 57th St, Chicago, IL 60637}
}

\date{\today}

\maketitle
\begin{abstract}
There is a well known analogy between the Laughlin trial wave function for the fractional quantum Hall effect, and the Boltzmann factor for the two-dimensional one-component plasma. The latter requires {{} analytic continuation} beyond the finite geometry used in its derivation. We consider both disk and cylinder geometry, and focus attention on the 
exact and asymptotic features of the edge density. At the special coupling $\Gamma := q^2/k_BT=2$ the system is exactly solvable. In particular the $k$-point correlation can be written as a $k \times k$ determinant, {{} allowing the edge density to be computed to first order in
$\Gamma - 2$. A double layer structure is found, which in turn implies an overshoot
of the density as the edge of the leading support is approached from inside the plasma.}
 Asymptotic analysis shows that the deviation from the leading order (step function) value is different for into the plasma than for outside. For general $\Gamma$, a Gaussian fluctuation formula is used to study the large deviation form of the density for $N$ large but finite. This asymptotic form involves thermodynamic quantities which we independently study, and moreover an appropriate scaling gives the asymptotic decay of the limiting edge density outside of the plasma.
\end{abstract}


\noindent 
{\it Dedicated to the memory of Bernard Jancovici
  (1930--2013) and his work on sum rules and exact solutions for
  Coulomb systems.}

\section{Introduction}
In the theory of the fractional quantum Hall effect the so-called Laughlin states are trial
wave functions in a two-dimensional domain of the form
\begin{equation}\label{IG1}
{1 \over C_N} \prod_{l=1}^N f(z_l) \prod_{1 \le j < k \le N} (u(z_k) - u(z_j))^m, \quad
z_l := x_l + i y_l.
\end{equation}
Here $m$ is even (odd) for bosonic (fermionic) states, and $m$ furthermore determines the filling
fraction $\nu$ of the lowest Landau level according to $\nu = 1/m$.

For planar geometry in the symmetric gauge
\begin{equation}\label{IG2}
f(z) = e^{-|z|^2/4 l_B^2}, \qquad u(z) = z,
\end{equation}
where $l_B = \sqrt{\hbar c/eB}$ is the magnetic length. For cylinder
geometry, with axis along the $y$-axis and perimeter $W$, in
the Landau gauge
\begin{equation}\label{IG3}
f(z) = e^{-y^2/2l_B^2}, \qquad u(z) = e^{2 \pi i z/W}.
\end{equation}
Below we set the units of length so that $l_B = 1/\sqrt{m}$. We refer to the wave function (\ref{IG1})
in the case (\ref{IG2}) by $\Psi_N^{\rm d}$, and in the case (\ref{IG3}) by $\Psi_N^{\rm c}$.

Our primary interest in this paper is in the particle density
\begin{equation}\label{PD}
\rho_{(1)}(z;m) := N \int_\Omega dx_2 dy_2 \cdots  \int_\Omega dx_N dy_N  \,
|\Psi_N(z,z_2,\dots,z_N) |^2.
\end{equation}
To leading order, in the planar geometry specified by (\ref{IG2}), $\rho_{(1)}(z;m) = 1/(2 \pi)
\chi_{|z| < \sqrt{N}}$, while in the cylinder geometry specified by  (\ref{IG3}),
$\rho_{(1)}(z;m) = 1/(2 \pi) \chi_{z \in \mathcal R}$, where $\mathcal R = \{ 0 \le x \le W,
\: 0 \ge y \ge - L \}$, $N/(WL) = 1/(2 \pi)$. Here $\chi_J = 1$ for $J$ true, and $\chi_J = 0$
otherwise. These behaviours are most easily seen by appealing to an interpretation of
$|\Psi_N|^2$ in terms of the Boltzmann factor for the classical two-dimensional 
one-component plasma; see Section \ref{Se1}. On the boundary of the leading support there
is a non-trivial double layer, or overshoot, behaviour characterized by a local maximum in the
density \cite{DMF96,CW03,MH86,TF99,ZW06,Wi12}, and it is our aim to undertake a study of
some of its analytic properties in the $N \to \infty$ limit.
It turns out that thermodynamic quantities of the plasma, such as the free energy and surface tension, appear in the associated asymptotic forms, so it is necessary to first undertake a study
of the thermodynamic properties of the plasma, which we do in Section \ref{SU}.

In particular, in Section \ref{SU} we pool together knowledge from previous studies to specify as
many terms as possible in the large $N$ expansion of the free energy. 
The coupling constant in the plasma is $\Gamma = q^2/k_B T$ (see below (\ref{eq:d3})).
In terms of quantities in (\ref{IG1}) we have $\Gamma = 2m$. Unlike $m$, the coupling $\Gamma$
is naturally a continuous variable.
The dependence on $\Gamma$ of the resulting
expressions are tested and illustrated by a combination of exact analytic, and exact numerical
results. In relation to exact analytic results, the case $\Gamma = 2$, which in the interpretation
(\ref{IG1}) corresponds to free fermions {{} in a magnetic field in the lowest Landau level}, is exactly solvable for both planar \cite{Gi65}
and cylinder \cite{CFS83} geometry. Knowledge of the exact one and two-point correlations
can be used to expand the free energy to first order in $\Gamma - 2$. And for $\Gamma = 4$,
6 and 8 expansion methods of the products of differences in
(\ref{IG1}) based on Jack polynomials (see Section \ref{SR}) can be used to provide exact
numerical data up to $N = 14$.

Our study of the edge density begins in Section \ref{S1}.
 Following the lead of the earlier work of Jancovici
\cite{Ja81} in the bulk, knowledge of the exact one, two, and three-point correlations
in the case of the planar geometry for $\Gamma = 2$ was recently used \cite{TF12} to 
calculate the exact form of the density to first order in $\Gamma - 2$. We
provide its $N \to \infty$ form in the case that the coordinates are centred on the 
boundary
of the leading support {{} for finite $N$}, and we show too that the same analytic expression results by computing
the edge scaling of the density computed to first order in $\Gamma - 2$ for cylinder geometry.
Moreover, the asymptotic  behaviour {{} into and outside the plasma}
can determined, and it is found the deviation
from the leading order (step function) value is different {{} in the two cases}.
The results of this section have been reported in a Letter by the present authors \cite{CFTW13},
which furthermore {{} casts them in the context of} the Laughlin droplet interpretation.

In Section \ref{SK} we study the large deviation form of the density outside of the leading
support, for $N$ large but finite. Our main tool here is to express (\ref{PD}) in terms of
the characteristic function for the distribution of a certain linear statistic, then to compute
its large $N$ form by using a Gaussian fluctuation formula. By an appropriate scaling of this
expression we obtain a prediction for the asymptotic decay of the edge density in the region
outside of the leading support for general $\Gamma > 0$.

\section{Plasma Viewpoint}\label{Se1}
The observation that the absolute value squared of the Laughlin trial wave functions for the fractional quantum Hall effect have an interpretation as the Boltzmann factor for certain two-dimensional one component plasma $(2d{\rm OCP})$ systems was already made in the original paper of Laughlin \cite{La83}. Generally the $2d{\rm OCP}$ refers to a system of $N$ mobile point particles of the same charge $q$ and a smeared out neutralising background, with the domain a two-dimensional surface. The charges interact via the solution $\Phi (\vec{r}, \vec{r}')$ of the Poisson equation on the surface. Thus for the plane
\begin{equation}
  \label{eq:d1}
  \Phi (\vec{r}, \vec{r}') = - \log (| \vec{r}- \vec{r}'|/l),
\end{equation}
where $l$ is an arbitrary length scale (we take $l = 1$), while for periodic bounday conditions in the $x$-direction, period $W$ (or equivalently a cylinder of circumference length $W$)
\begin{equation}
  \label{eq:d2}
  \Phi (\vec{r}, \vec{r}') = - \log \Big(| \sin \big( \pi (x-x' + i(y - y'))/W \big) \big(\frac{W}{\pi} \big) \Big),
\end{equation}

With $\beta := 1/k_B T$ the Boltzmann factor for a classical system is $e^{-\beta U}$, where $U$ is the total potential energy. As detailed in \cite[\S1.4.1]{Fo10}, $U = U_1+U_2+U_3$, where $U_1$ corresponds to the particle-particle interaction, $U_2$ to the particle-background interaction, and $U_3$ to the background-background interaction. In the case that the domain is a plane, with the smeared out neutralizing background a disk  at the origin of radius $R$, the particles couple to the background via a harmonic potential towards the origin. 
{{} Explicitly one has 
\begin{equation}\label{u1}
U_1 = - \sum_{1 \le j < k \le N} \log | \vec{r}_k - \vec{r}_j |
\end{equation}
and
\begin{equation}\label{u2}
U_2 + U_3 =  N^2 \Big ( {1 \over 4} \log N - {3 \over 8} \Big )
+ {1 \over 2} \ \sum_{j=1}^N \vec{r}_j^2 ,
\end{equation}}
and so the explicit form of the Boltzmann factor is  (see e.g.~\cite[eq. (1.72)]{Fo10})
\begin{equation}
  \label{eq:d3}
  A_{N,\Gamma}^{\rm d} e^{-\pi \Gamma \rho_b \sum_{j=1}^N |\vec{r_j}|^2/2}\prod_{1 \leq j < k \leq N}|\vec{r_k} - \vec{r_j}|^\Gamma, \qquad A_{N,\Gamma}^{\rm d} = e^{-\Gamma N^2 (\frac{1}{2} \log R - \frac{3}{8})}
\end{equation}
where $\rho_b = N/\pi R^2$ and $\Gamma = q^2/k_B T$.
The derivation of (\ref{eq:d3}) requires the particles be confined to the disk of the smeared out background and thus $|\vec{r_j}| \leq R$. To get an analogy with the absolute value squared of the trial wave functions (\ref{IG1}) we must relax this condition by allowing the domain to be all of $\mathbb{R}^2$; this will be referred to as soft disk geometry. 

With $Z_{N,\Gamma}^{\rm d}$ denoting the partition function corresponding to (\ref{eq:d3}), i.e.~(\ref{eq:d3}) integrated over $\vec{r}_j \in \mathbb R^2$ $(j=1,\dots,N)$ and multiplied by $1/N!$, one has that for $\Gamma = 2$ (see e.g.~\cite[above eq.~(3.14)]{TF99})
$$
Z_{N,2}^{\rm d} = \pi^N e^{3 N^2 / 4} N^{-N^2/2} (\pi \rho_b)^{-N/2} G(N+1),
$$
where $G(N+1) := \prod_{l=1}^{N-1} l!$. Consequently \cite[eq.~(3.14)]{TF99}
$$
\beta F_{N,2}^{\rm d} := - \log Z_{N,2}^{\rm d} =N \beta f(2,\rho_b) + {1 \over 12} \log N - \zeta'(-1) + \mathcal O\Big ( {1 \over N^2} \Big ),
$$
where
\begin{equation}\label{F2a}
 \beta f(2,\rho_b) = {1 \over 2} \log \Big ( {\rho_b \over 2 \pi^2} \Big ).
\end{equation}
Furthermore, the one-body density can similarly be computed exactly at $\Gamma = 2$ with the result {{} (see e.g.~\cite[Prop.~15.3.4]{Fo10})}
\begin{equation}\label{eq:C2}
 \rho_{(1)}^{\rm d} (\vec{r};2) = {1 \over \pi} {\Gamma(N;r^2) \over \Gamma(N)}.
\end{equation} 

In the case of semi-periodic boundary conditions, the neutralizing background is chosen to be the rectangle $0 < x < W$, $0 < y < L$, and the particles couple to the background via a harmonic potential in the $y$-direction only, centred at $y = L/2$  \cite{CFS83}. For the corresponding Boltzmann factor we find
\begin{equation}
  \label{eq:d4}
  A_{N,\Gamma}^{\rm c} e^{-\pi \Gamma \rho_b \sum_{j=1}^N (y_j -L/2)^2}\prod_{1 \leq j < k \leq N} \big |2 \sin \frac{\pi (x_j - x_k + i(y_j -y_k))}{W} \big |^\Gamma,
\end{equation}
where
\begin{equation}
  \label{eq:d5}
  A_{N,\Gamma}^{\rm c} = \big(\frac{W}{2\pi} \big)^{-N\Gamma /2} e^{- \frac{\pi \Gamma}{12} NL^2 \rho_b}
\end{equation}
and $\rho_b = N/LW$. Analogous to the situation with (\ref{eq:d3}), the derivation of (\ref{eq:d4}) requires $0 < y_j < L$, but to get an analogy with the absolute value squared of the trial wave function (\ref{IG1}) in the case (\ref{IG3}) we must relax this conditions, obtaining what will be referred to as soft cylinder geometry. 
For $\Gamma = 2$, results from \cite{CFS83} tell us that
\begin{equation}\label{F1}
{1 \over N} \beta F_{N,2}^{\rm c}  =   \beta f(2,\rho_b) + {\pi \rho_b L^2 \over 6 N^2}
\end{equation}
and
\begin{equation}\label{F12}
 \rho_{(1)}^{\rm c} (\vec{r};2) = {1 \over W} \sqrt{2 \rho_b} \sum_{m=0}^{N-1} \exp \Big ( - 2 \pi \rho_b \Big ( y - {m + 1/2 \over W \rho_b} \Big )^2 \Big ).
 \end{equation}
 
 \section{Universal properties of the free energy}\label{SU}
 \subsection{Introductory remarks}
 Consider first the soft disk geometry. For general $\Gamma > 0$ one expects the large $N$ expansion
 \begin{equation}\label{k1}
 \beta F_{N,\Gamma}^{\rm d} := -  \log Z_{N,\Gamma}^{\rm d} =N \beta f(\Gamma,\rho_b)  + \beta \mu(\Gamma,\rho_b) (2 \pi \sqrt{N/(\pi \rho_b)}) +
 {1 \over 12} \log N + \mathcal O(1).
 \end{equation}
 In the leading term, $ \beta f(\Gamma,\rho_b) $ is the dimensionless free energy per particle.
 The universal term ${1 \over 12} \log N$ was identified by relating the plasma to a free Gaussian field \cite{JMP94}.
 An unpublished result of Lutsyshin makes the conjecture
  \begin{equation}\label{k2}
  \beta \mu(\Gamma,\rho_b) = {\sqrt{\pi \rho_b} \over 2 \pi} {4 \log (\Gamma/2) \over 3 \sqrt{\pi}} =  \sqrt{\rho_b} {2 \log (\Gamma/2) \over 3 \pi} .
  \end{equation}
  Since the radius of the background is $2 \pi \sqrt{N/(\pi \rho_b)}$, $ \mu(\Gamma,\rho_b)$ has the interpretation as a  surface tension. Note that
  (\ref{k2}) is consistent with the exact result (\ref{F2a}) as it gives $ \beta \mu(2,\rho_b) = 0$.
  
  Consider now the soft cylinder. Universality of the  dimensionless free energy per particle and the surface tension
  imply that for large $N$
  \begin{equation}\label{k3}
 \beta F_{N,\Gamma}^{\rm c} := -  \log Z_{N,\Gamma}^{\rm c} =N \beta f(\Gamma,\rho_b)  + \beta \mu(\Gamma,\rho_b) (2W) +
 {\pi \rho_b L^2 \over 6 N} +  \mathcal O(1).
  \end{equation} 
 Here the universal term $\pi \rho_b L^2 /6 N^2$ is a consequence of the relationship between the plasma on {{} an infinitely long} cylinder and the corresponding
 Gaussian free field \cite{Fo91}.
 
 \subsection{Validity of free energy expansion for $\Gamma = 2 + \epsilon$ $(\epsilon \ll 1)$}
Consider the soft disk plasma system with mobile particles having charge $q=1$ and total energy $U$  (recall \S \ref{Se1}). It follows from the definitions
that to first order in $\Gamma - 2$,
  \begin{equation}\label{k6} 
 \beta F_{N,\Gamma}^{\rm d} -  \beta F_{N,2}^{\rm d} =  (\Gamma - 2) \langle U \rangle \Big |_{\Gamma = 2},
\end{equation}
{{} where $U$ denotes the total energy. But we know from above 
 that $U = U_1 + U_2 + U_3$, with $U_1$  the potential energy of the 
 particle-particle interaction as given by (\ref{u1}), and $U_2+U_3$ the sum of the
 particle-background and background-background interactions as given by (\ref{u2}).} A result of Shakirov \cite{Sh11} tells us that
    \begin{equation}\label{k4} 
  \langle U_1 \rangle \Big |_{\Gamma = 2 \atop \rho_b = 1/\pi} =
  - {1 \over 2} \Big ( {N^2 \over 2} \log N - {N^2 \over 4} + {1 \over 2} (1 + {\bf C} )N - {4 \over 3} \sqrt{N \over \pi} + {5 \over 24} +  \mathcal O({1 \over \sqrt{N}}) \Big ),
 \end{equation} 
  where $\bf C$ denotes Euler's constant. The remaining averages are simple to compute.
  
  \begin{lemma}\label{Lb1}
  We have
  $$
  \langle U_2 + U_3 \rangle \Big |_{\Gamma = 2 \atop \rho_b = 1/\pi} =   {N^2 \over 4} \log N - {N^2 \over 8} + {N \over 4}.
  $$
  \end{lemma}
  
  \noindent
  Proof. \quad We see from (\ref{u2}) that
   \begin{equation}\label{k4a} 
  \langle U_2 + U_3 \rangle \Big |_{\Gamma = 2 \atop \rho_b = 1/\pi} = N^2 \Big ( {1 \over 4} \log N - {3 \over 8} \Big ) +
  {1 \over 2} \Big \langle \sum_{j=1}^N \vec{r}_j^{\, 2} \Big \rangle     \Big |_{\Gamma = 2 \atop \rho_b = 1/\pi}.
  \end{equation} 
  Introducing the configuration integral
 \begin{equation}\label{Q1}
  Q_{N,\Gamma}^{\rm d} (\rho_b) := \int_{{\mathbb R}^2}d\vec{r}_1 \cdots \int_{{\mathbb R}^2}d\vec{r}_N e^{-(\pi \rho_b \Gamma /2)\sum_{j=1}^N \vec{r}_j^2}\prod_{1 \leq j < k \leq N}|\vec{r}_k - \vec{r}_j|^{\Gamma},
\end{equation}
we see that
$$
 \Big \langle \sum_{j=1}^N \vec{r}_j^{\, 2} \Big \rangle     \Big |_{\Gamma = 2 \atop \rho_b = 1/\pi} = - {1 \over \pi}
{\partial \log Q_{N,2}^{\rm d}(\rho_b) \over  \partial \rho_b}   \Big |_{\rho_b = 1/\pi}.
$$
On the other hand, a simple scaling shows
$$
 Q_{N,\Gamma}^{\rm d}(\rho_b) = \rho_b^{-\Gamma N (N-1)/4 - N} Q_{N,\Gamma}^{\rm d}(1),
 $$
 so we obtain {{}
 \begin{equation}
 \Big \langle \sum_{j=1}^N \vec{r}_j^{\, 2} \Big \rangle     =   {1 \over \pi \rho_b}
 \Big ( 
 {\Gamma N (N-1) \over 4} + N \Big )
 \end{equation}}
 Setting $\Gamma = 2$, $\rho_b = 1/\pi$ and substituting in (\ref{k4a}) gives the stated result. \hfill $\square$
 
 \medskip
 Adding (\ref{k4}) to the result of Lemma \ref{Lb1} and substituting in (\ref{k6}) we have to first order in $\Gamma - 2$
   \begin{equation}\label{k7}  
( \beta F_{N,\Gamma}^{\rm d} -  \beta F_{N,2}^{\rm d}  )\Big |_{\rho_b=1/\pi}=  (\Gamma - 2) \Big ( - {{\bf C} N   \over 4} + {2 \over 3} \sqrt{N \over \pi}
 - {5 \over 48} + \mathcal O \Big ( {1 \over \sqrt{N}} \Big ) \Big ).
 \end{equation}
 In particular, the terms proportional to $\sqrt{N}$ is in precise agreement with the conjecture (\ref{k2}) expanded to the same order. As an aside, we
 remark that (\ref{k7}) and (\ref{k6}) together tell us that to leading order in $N$, $\langle U \rangle$ with $\Gamma = 2$ and $\rho_b = 1/\pi$, is
 equal to $- {\bf C}N/4$. This is a result first deduced by Jancovici \cite{Ja81}, {{}
 using the relationship of the leading form of $\langle U \rangle$ and an average of
 the potential $-\log |\vec{r}|$ with respect to the bulk truncated two-point function.}
 
 The formula (\ref{k6}) also applies with the soft disk replaced by the soft cylinder; however the analogue of (\ref{k4}) is not in the existing literature.
 Making use of knowledge of the exact form of the one and two-point functions for the soft cylinder geometry at $\Gamma = 2$ \cite{CFS83} we
 find (see Appendix \ref{AA1})
 \begin{equation}\label{V1}
   \langle U_1 \rangle^{\rm c} \Big |_{\Gamma = 2} = {1 \over 2} \Big (- {\pi \over W^2 \rho_b} {N^3 \over 3} - N \log \Big ( \sqrt{\rho_b \over \pi} {W \over 2} \Big )
   - {N \over 2} - {N {\bf C} \over 2} + {4 \sqrt{\rho_b} \over 3 \pi} W  \Big )+ \mathcal O(1).
   \end{equation}
Furthermore, a more elementary computation, making only use of the one-point function (\ref{F12}), gives
 \begin{equation}\label{V2}
  \langle U_2 + U_3 \rangle^{\rm c} \Big |_{\Gamma = 2}  = {N \over 2} \log {W \over 2 \pi} + {\pi \over 6} N L^2 \rho_b + {N \over 4}.
  \end{equation}
  Thus to first order in $\Gamma - 2$,
  \begin{equation}\label{V3}
 \beta F_{N,\Gamma}^{\rm c} -  \beta F_{N,2}^{\rm c}  =  (\Gamma - 2) \Big ( - {N \over 2} \log \sqrt{\pi \rho_b} - {{\bf C} N \over 4} +
 {2 \sqrt{\rho_b} \over 3 \pi} W + \mathcal O(1) \Big ).
 \end{equation} 
 In particular, the term proportional to $W$ is consistent with the expansion (\ref{k3}). 
 
 \subsection{Exact numerical results {{} for the free energy at} $\Gamma =$ 4, 6 and 8}\label{SR}
 Let $\Gamma = 4p$, $p \in \mathbb Z^+$, and let $\mu = (\mu_1,\dots,\mu_N)$ be a partition of $pN(N-1)$ such that
$$
2p (N - 1) \ge \mu_1 \ge \mu_2 \ge \cdots \ge \mu_N \ge 0.
$$
Also, let $m_i$ denote the corresponding frequency of the integer $i$ in $\mu$, let $S_N$ denote the set of permutations of $N$,
and define the corresponding monomial symmetric function by
$$
m_\mu(z_1,\dots,z_N) = {1 \over \prod_i m_i!} \sum_{\sigma \in S_N} z_{\sigma(1)}^{\mu_1} \cdots  z_{\sigma(N)}^{\mu_N}.
$$
A method based on symmetric Jack polynomials  \cite{BH08a} gives, for small $p$, an efficient way to compute the
coefficients $\{ c_\mu^{(N)}(2p) \}$
$$
\prod_{1 \le j < k \le N} (z_k - z_j)^{2p} = \sum_\mu c_\mu^{(N)}(2p) m_\mu(z_1,\dots,z_N) .
$$
This is significant since then we have \cite{TF99}
\begin{equation}\label{k8}
  Q_{N,4p}^{\rm d} (1/\pi) = N! \pi^N \sum_\mu {(c_\mu^{(N)}(2p))^2 \over \prod_i m_i!} \prod_{l=1}^N \mu_i!
\end{equation}

Similar considerations hold true for $\Gamma = 4p+2$. Now we must take $\mu$ to be a partition of $(p+1)N(N-1)$ such that
$$
(2p+1)(N-1) \ge \mu_1 > \mu_2 > \cdots > \mu_N \ge 0.
$$
With $s_\nu(z_1,\dots,z_N)$ denoting the Schur polynomials, we then use the anti-symmetric Jack polynomials to expand \cite{BR09}
$$
\prod_{1 \le j < k \le N} (z_k - z_j)^{2p} = \sum_\mu c_\mu^{(N)}(2p+1) s_{\mu-\delta_N}(z_1,\dots,z_N) ,
$$
where $\delta_N := (N-1,N_2,\dots,0)$. Consequently \cite{TF99}
\begin{equation}\label{k9}
  Q_{N,4p+2}^{\rm d} (1/\pi) = N! \pi^N \sum_\mu (c_\mu^{(N)}(2p+1))^2 \prod_{l=1}^N \mu_i!
\end{equation}
Using (\ref{k8}) and (\ref{k9}), we computed numerically the free
energy in the soft disk for $\Gamma=4$ and $6$ with $N$ ranging from
$2$ to $14$, and for $\Gamma=8$ with $N=2$ to 11. In order to test the
expansion (\ref{k1}), the data for $N=12,13,14$ ($\Gamma=4, 6$) and
$N=9,10,11$ ($\Gamma=8$) is fitted to the ansatz
\begin{equation}
  \label{eq:Ffitdisk}
  \beta F_{N\Gamma}^{\rm d} = N\beta f(\rho_b,\Gamma) 
    + \beta \mu(\Gamma,\rho_b) (2 \pi \sqrt{N/(\pi \rho_b)}) +
    {1 \over 12}  \log N + d
\end{equation}
The data obtained for $g(\Gamma)=\beta
f(\Gamma,\rho_b)-\left(1-\frac{\Gamma}{4}\right)\log \rho_b$,
$\beta\mu(\Gamma,\rho_b)$ and $d$ is shown in Table~\ref{tabdisk}. The
results for the bulk free energy $\beta f$ reproduces known numerical
estimates obtained by studying the 2dOCP in a sphere for $\Gamma=4,
6$~\cite{TF99} and 8 \cite{T13} within a very small margin of error:
less than 0.02\% for $\Gamma=4$ and 6, and 0.16\% in the worst case
$\Gamma=8$. The surface tension term $\beta \mu$ is compared with the
conjecture (\ref{k2}) and the results give a strong support to this
conjecture as they only differ by less than 1\% for $\Gamma=4$ and 6,
and 5.5\% for $\Gamma=8$.
\begin{table}[tbp]
  \centering
  \begin{tabular}{|c||c|c|c|}
    \hline
    $\Gamma$ & 4 & 6 & 8 \\
    \hline 
    $g$ (soft disk) & -2.44972  & -3.51707 & -4.64639 \\
    $g$ (sphere)   & -2.449884 & -3.5175 & -4.639 \\
    Relative 
    difference   & 0.007\%   & 0.012\% & 0.16\% \\
    \hline
    $\beta \mu /\sqrt{\rho_b}$ & 0.145938 & 0.232798  & 0.310371
    \\
    Conjecture (\ref{k2}) & 0.147090 & 0.233132
    & 0.29418
    \\
    Relative difference & 0.78 \% & 0.14\% & 5.5\%
    \\
    \hline
    $d$ & -0.0244379 & -0.163993 & -0.353555
    \\
    \hline
  \end{tabular}
  \caption{Fitting the free energy in the soft disk as specified in (\ref{eq:Ffitdisk}).}
  \label{tabdisk}
\end{table}

A more extensive numerical study can be done in the soft cylinder
geometry as $W$ and $N$ can be varied independently, and more
numerical data can be obtained for the free energy.
Formulas analogous to (\ref{k8}) and (\ref{k9}) hold true for the soft cylinder. There the relevant configuration integral is
\begin{align}\label{Q2}
  Q_{N, \Gamma}^{\rm c} (L, W)  = &  \int_{-\8}^\8 dy_1 \cdots \int_{-\8}^\8 dy_N e^{-\Gamma \pi \rho_b \sum_{l=1}^N (y_l - L/2)^2} \int_0^Ldx_1 \cdots \int_0^L dx_N \nonumber \\
& \times \prod_{1 \leq j < k \leq N}\Big|2\sin \frac{\pi (x_k - x_j)+i\pi(y_k - y_j)}{W} \Big|^\Gamma .
\end{align}
For $\Gamma$ even and $w_j := x_j + i y_j$ we have
$$
|2 \sin \pi (w_k - w_j) |^\Gamma = e^{\pi (y_j + y_k)\Gamma/2}
( e^{2 \pi i w_j} - e^{2 \pi i w_k})^{\Gamma/2} ( e^{-2 \pi i \bar{w}_j} - e^{-2 \pi i \bar{w}_k})^{\Gamma/2}.
$$
Consideration of the derivation leading to (\ref{k8}) and (\ref{k9}) we then have 
\begin{equation}\label{k10}
  Q_{N, 4p}^{\rm c} (\sqrt{N}, \sqrt{N}) = N^{N/2} N! \Gamma^{-N/2}
  e^{-\pi \Gamma N^2/4} \sum_\mu {(c_\mu^{(N)}(2p))^2 \over \prod_i m_i!} e^{\pi \Gamma \sum_{j=1}^N (2\mu_j/\Gamma + 1/2)^2/N}
\end{equation}
and
\begin{equation}\label{k10b}
  Q_{N, 4p+2}^{\rm c} (\sqrt{N}, \sqrt{N}) = N^{N/2} N! \Gamma^{-N/2}
  e^{-\pi \Gamma N^2/4} \sum_\mu (c_\mu^{(N)}(2p+1))^2 e^{\pi \Gamma \sum_{j=1}^N (2\mu_j/\Gamma + 1/2)^2/N}.
\end{equation}

More generally, if $W$ is considered as an independent variable from
$N$, let us define $\lambda=(\rho_b W)^{-1}$ which is a characteristic
length of the problem: as shown in \cite{CFS83, SWK04} the one-body
density is periodic along the $y$-axis with period $\lambda$ when
$N\to\infty$ and $W$ fixed. Let $\tilde{W}=W/\lambda=\rho_b W^2$ be
the cylinder circumference scaled out by $\lambda$. The configuration
integral~(\ref{Q2}) is
\begin{equation}\label{k10c}
  Q_{N, \Gamma}^{\rm c} (L, W) = \rho_b^{-N}\tilde{W}^{N/2} N! \Gamma^{-N/2}
  e^{-\pi \Gamma N^3/(4\tilde{W})} \sum_\mu {(c_\mu^{(N)}(\Gamma/2))^2 \over \prod_i m_i!} e^{\pi \Gamma \sum_{j=1}^N (2\mu_j/\Gamma + 1/2)^2/\tilde{W}}\,,
\end{equation}
valid for both cases $\Gamma=4p$ and $\Gamma=4p+2$. In the latter case
$\prod_im_i!=1$ as in all the partitions $\mu$ with
$c_{\mu}^{(N)}(2p+1)\neq0$ all frequencies are $m_i=1$. The free energy
is given by
\begin{eqnarray}
  \beta F_{N,\Gamma}^{c}(\tilde{W})& =& N\left(1-\frac{\Gamma}{4}\right)\log \rho_b 
  - \frac{N}{2}\left(1-\frac{\Gamma}{2}\right)\log \tilde{W} 
  +\frac{N}{2}\log\Gamma - \frac{N\Gamma}{2}\log (2\pi) 
  \nonumber\\
  &&   + \frac{\pi\Gamma N^3}{3\tilde{W}} -\log Q_{N,\Gamma}^{\rm c *}(\tilde{W})
  \label{eq:Fcyl}
\end{eqnarray}
with
\begin{equation}
\label{eq:Qstar-cyl}
Q_{N,\Gamma}^{\rm c *}(\tilde{W})=
\sum_\mu {(c_\mu^{(N)}(\Gamma/2))^2 \over \prod_i m_i!} e^{\pi \Gamma \sum_{j=1}^N (2\mu_j/\Gamma + 1/2)^2/\tilde{W}}
\,.
\end{equation}

\begin{figure}[tbp]
  \centering
  \includegraphics[width=8cm]{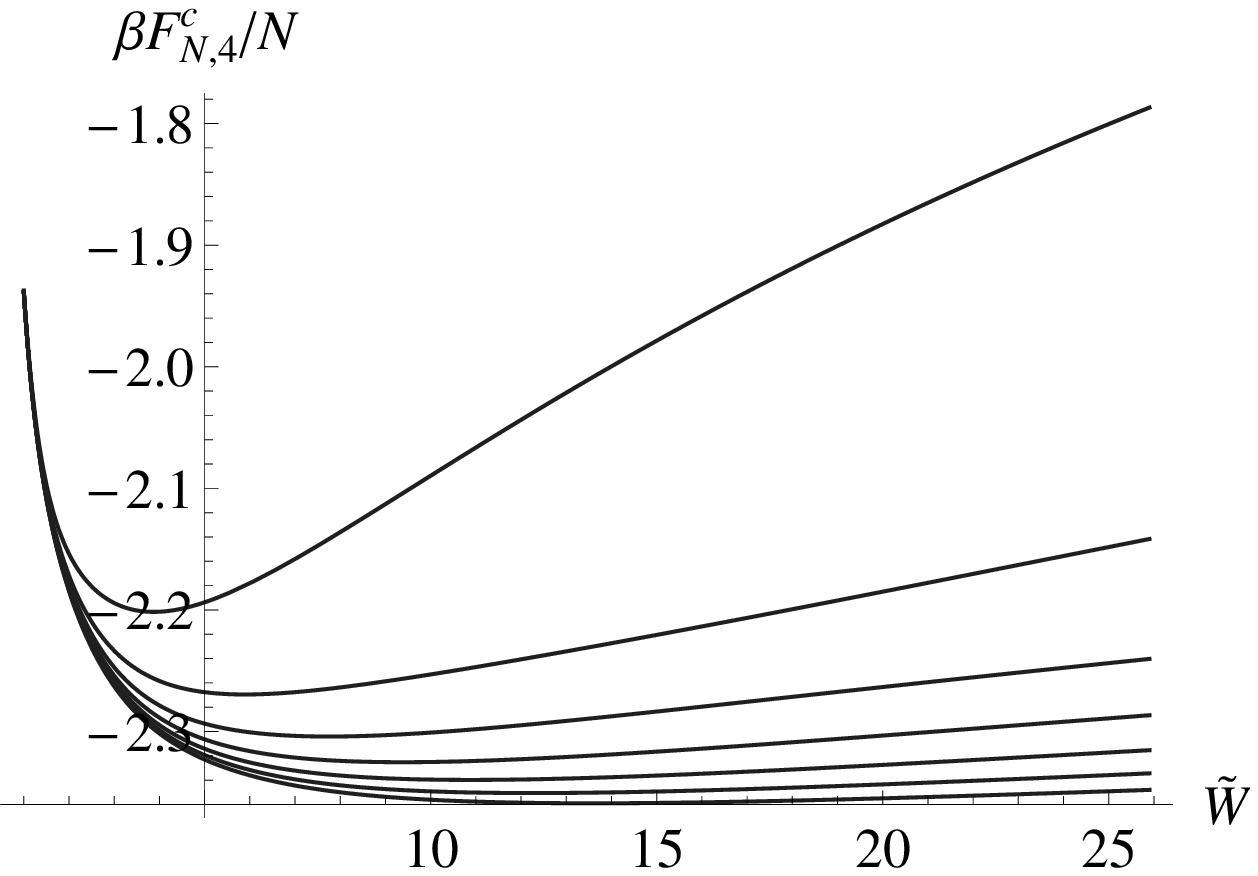}
  \includegraphics[width=8cm]{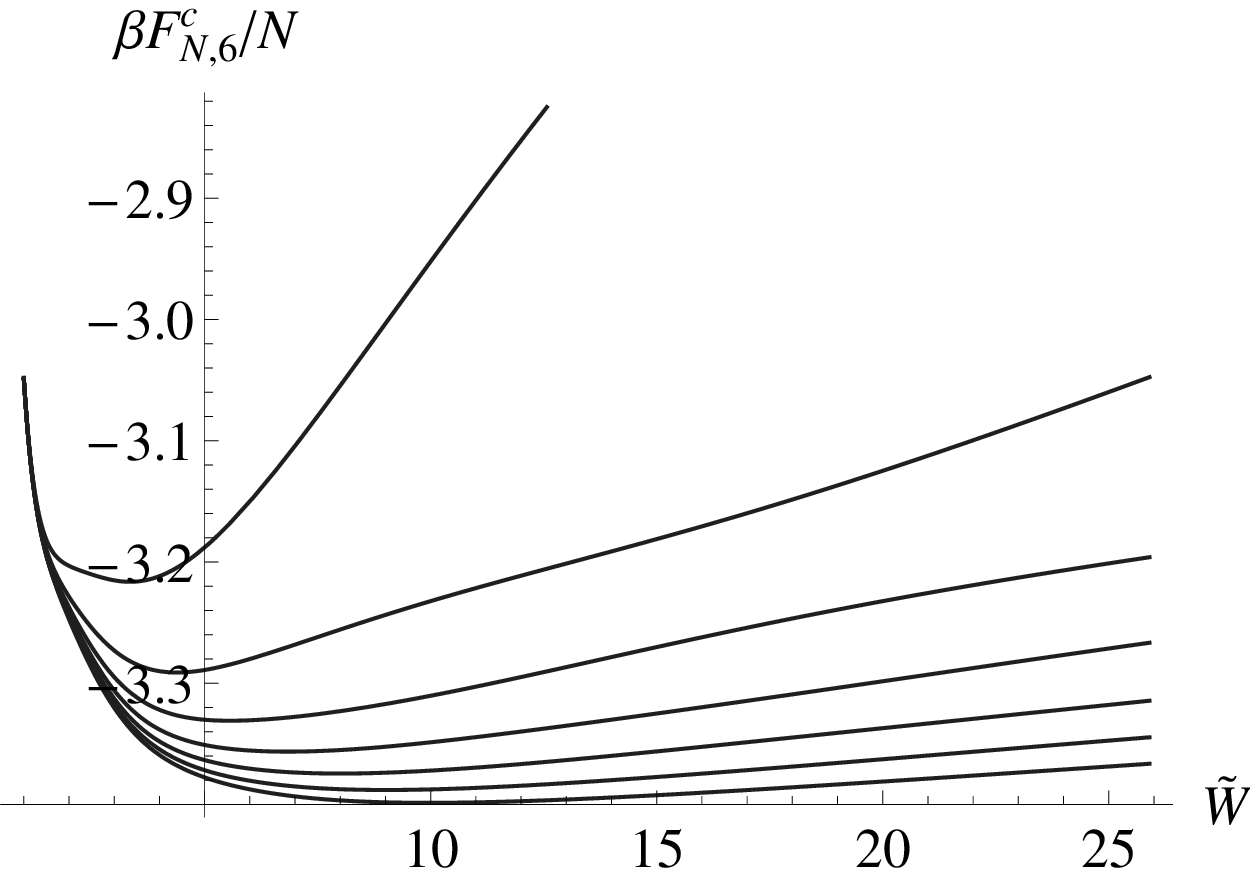}

  \includegraphics[width=8cm]{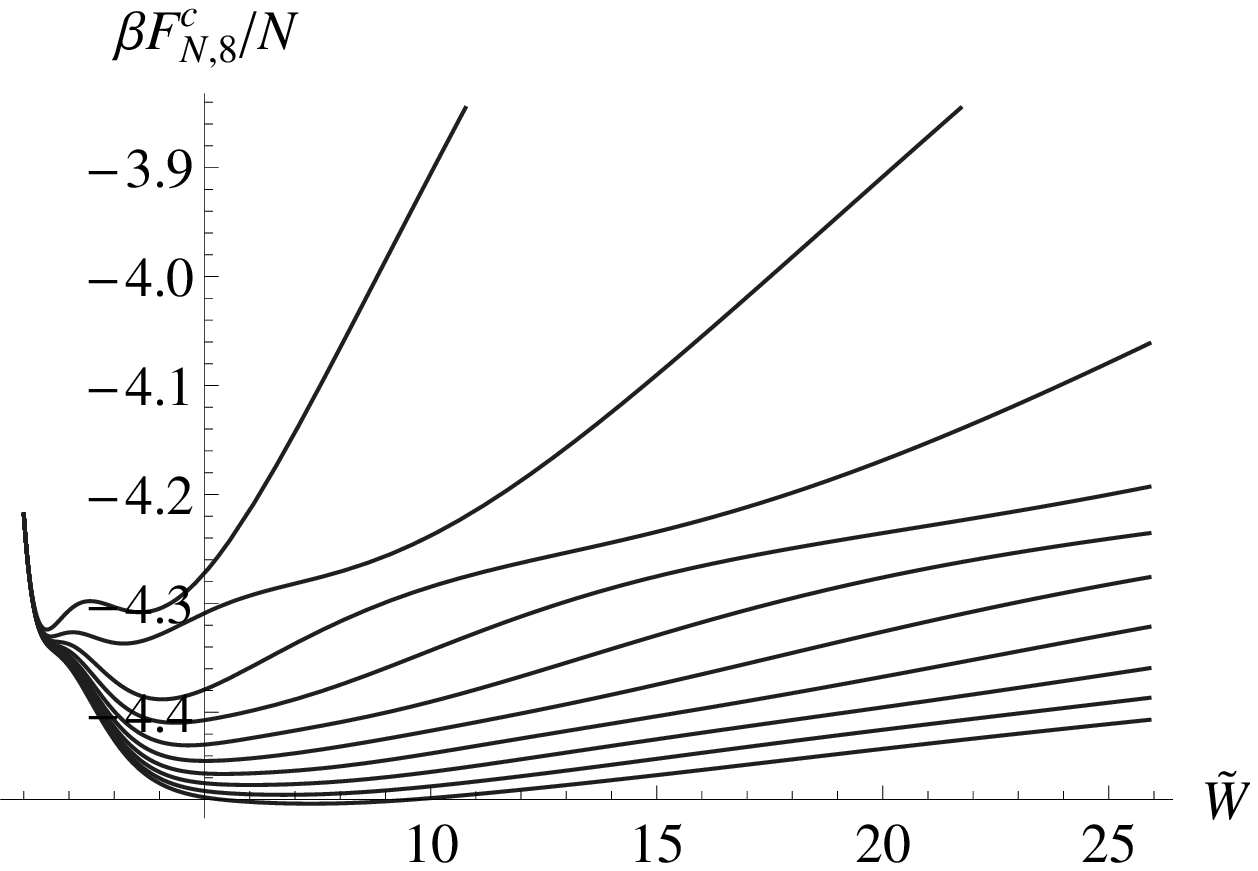}
  \caption{Soft cylinder free energy as a function of the cylinder
    circumference $\tilde{W}$ when $\Gamma=4$ (top left), $\Gamma=6$
    (top right) and $\Gamma=8$ (bottom). In each graph, from top to
    bottom, the number of particles is $N=2,4,6,8,10,12,14$ for
    $\Gamma=4$ and 6, and $N=2,3,4,5,6,7,8,9,10,11$ for $\Gamma=8$. 
{{}
These figures show that the free energy is minimum for a particular
value of the circumference $\tilde{W}$.
}
}
  \label{graF}
\end{figure}

\begin{figure}[tbp]
  \centering
  \includegraphics[width=10cm]{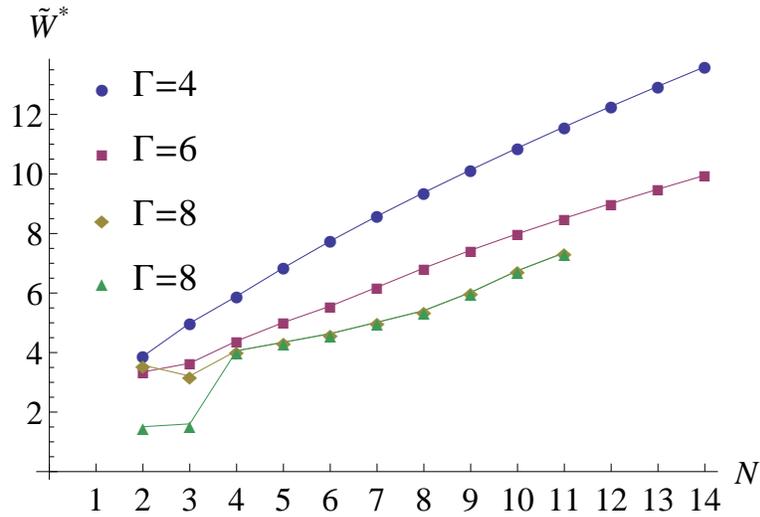}
  \caption{Location of the soft cylinder free energy minimum.}
  \label{gramini}
\end{figure}

We computed (\ref{eq:Fcyl}) numerically. The
calculations are computationally intensive for large values of $N$
because of the immense number of partitions involved, thus we had to
limit our efforts to $N$ varying from 2 to 14 for $\Gamma=4$ and
$\Gamma=6$, and $N$ from 2 to 11 for $\Gamma=8$. However, $\tilde{W}$
can be arbitrarily choosen without any computational increase
effort. We choose $\tilde{W}$ varying from 1 to 25.9 by increments of
0.1, {{} therefore exploring two different types of geometries:
  thin cylinder (small $\tilde{W}$) and thick cylinder (large $\tilde{W}$).}
The free energy is shown in Figure~\ref{graF} as a function of
$\tilde{W}$ for various values of $N$. For $\Gamma=4$ and 6, the free
energy exhibits a unique minimum for a particular value of
$\tilde{W}=\tilde{W}^*$ which depends on $N$. This is also the case
for $\Gamma=8$ and $N\geq 4$, however for $N=2$ and 3, the free energy
exhibits two minimums. In Figure~\ref{gramini}, the location of the
minimum $\tilde{W}^*$ is shown as a function of $N$. As $N$ increases,
also does $\tilde{W}^*$. The figure shows that, in the range of values
of $N$ considered, $\tilde{W}^*$ is of the same order of magnitude
that $N$, that is $W^*\propto \sqrt{N}$. {{} For large $N$,
  this corresponds to thick cylinders, thus suggesting that at a given
  density, thick cylinders are more stable thermodynamically than
  thin cylinders. In the following sections we will be interested in
  the scaling laws for thick cylinders.}

As the free energy expansion (\ref{k3}) is expected to hold for large
$N$ and large $W$, we sought to fit the numerical data corresponding
to $N>7$ and $\tilde{W}>7$ to an ansatz compatible with (\ref{k3}) of
the form
\begin{equation}
  \label{eq:Fexpans}
  \beta F_{N,\Gamma}^{\rm c} = N \beta f(\Gamma,\rho_b) + 2\beta \mu(\Gamma,\rho_b) \sqrt{\tilde{W}/\rho_b}
+ c\, \frac{N}{\tilde{W}} + d.
\end{equation}
The results for $g(\Gamma)=\beta
f(\Gamma,\rho_b)-\left(1-\frac{\Gamma}{4}\right)\log \rho_b$,
$\beta\mu(\Gamma,\rho_b)$, $c$ and $d$ are shown in Table
\ref{tabcyl}. The bulk free energy $\beta f$ is compared to the
numerical estimates obtained by studying the 2dOCP in a sphere for
$\Gamma=4, 6$~\cite{TF99} and 8 \cite{T13}. As it should be, the
difference is very small, less than 0.05\%. Also the universal
correction $c$ differs from the expected value $\pi/6$ only by less
than 4\% in the worst case ($\Gamma=8$). The numerical data again
strongly supports Lutsyshin's conjecture~(\ref{k2}) for the surface
tension term $\beta \mu$, as the relative difference between the
conjecture and the numerical data is less than 3\% in the worst case
($\Gamma=8$).

\begin{table}[tbp]
  \centering
  \begin{tabular}{|c||c|c|c|}
    \hline
    $\Gamma$ & 4 & 6 & 8 \\
    \hline 
    $g$ (cylinder) & -2.45003  & -3.5180 & -4.641 \\
    $g$ (sphere)   & -2.449884 & -3.5175 & -4.639 \\
    Relative 
    difference   & 0.006\%   & 0.014\% & 0.043\% \\
    \hline
    $\beta \mu /\sqrt{\rho_b}$ & 0.146534 & 0.235029 & 0.30112 
    \\
    Conjecture (\ref{k2}) & 0.147090 & 0.233132
    & 0.29418
    \\
    Relative difference & 0.378 \% & 0.814\% & 2.36\%
    \\
    \hline
    $c$ & 0.525251 & 0.529638 & 0.544192 \\
    Relative difference from $\pi/6$ 
     & 0.316\% & 1.15\% & 3.93\%     
    \\
    \hline
    $d$ & -0.347216 & -0.556913 & -0.715804
    \\
    \hline
  \end{tabular}
  \caption{Fitting the free energy in the soft cylinder as specified in (\ref{eq:Fexpans}).}
  \label{tabcyl}
\end{table}

In the next section, we will be interested in the scaled edge density
where $\tilde{W}=N\to\infty$. Notice that in that limit, the universal
correction to the free energy, $(\pi/6)\,\tilde{W}/N$, and the
$\mathcal O(1)$ correction in (\ref{k3}) ($d$ in (\ref{eq:Fexpans}))
become of the same order, and give a $\mathcal O(1)$ correction to the
free energy equal to $d+(\pi/6)$.


\section{Exact first order correction to the scaled edge density {{} at $\Gamma = 2$} }
\label{S1}
\subsection{Disk geometry}
In disk geometry, the density expanded about $\Gamma = 2$ has been computed to first order in $(\Gamma - 2)$ by T\'ellez and Forrester \cite{TF12}. To present their result, introduce
\begin{equation*}
  I(k_1, k_2) = \iint_{0 \leq t_2 < t_1} e^{-t_1-t_2}t_1^{k_1}t_2^{k_2}dt_1dt_2
\end{equation*}
\begin{align}
\label{eq:JI}
  J(k_1, k_2) & = \int_0^{\infty}\int_0^{\infty} e^{-t_1-t_2}t_1^{k_1}t_2^{k_2} \log \big({\rm max}(t_1,t_2)\big)dt_1dt_2 \nonumber \\
& = \frac{\partial}{\partial k_1} I(k_1, k_2) + \frac{\partial}{\partial k_2} I(k_2, k_1)
\end{align}
and let $\Gamma (k,x), \ \gamma(k,x)$ denote the usual upper and lower incomplete gamma functions. The result of \cite{TF12} reads
\begin{align}
  \label{eq:C1}
  \rho_{(1)}^{\rm d} (\vec{r};\Gamma) = & \rho_{(1)}^{\rm d} (\vec{r};2)\nonumber \\
& - \frac{(\Gamma -2)}{\pi} e^{-|z|^2}\Bigg \{\sum_{k_1=0}^{N-1} \sum_{\substack{k_2=0 \\ k_2 \ne k_1} }^{N-1} 
\frac{|z|^{2k_2}}{2k_1!(k_2!)^2}J(k_1,k_2)\nonumber \\
& + \sum_{k_1=0}^{N-1} \sum_{k_2=k_1+1}^{N-1} \frac{I(k_1, k_2)}{k_1!k_2!(k_2-k_1)} \bigg(\frac{|z|^{2k_2}}{k_2!} + \frac{|z|^{2k_1}}{k_1!}  \bigg) \nonumber \\
& - \sum_{k_1=0}^{N-1} \sum_{k_2=k_1+1}^{N-1} \frac{|z|^{2k_1}\gamma(k_2+1, |z|^2)+|z|^{2k_2}\Gamma(k_1+1, |z|^2)}{(k_2-k_1) k_1!k_2!} \nonumber \\
& - \sum_{k_2=0}^{N-1}\frac{|z|^{2k_2}}{2k_2!} \sum_{\substack{k_1=0 \\ k_1 \neq k_2 }}^{N-1} \frac{\gamma(k_1+1, |z|^2)\log(|z|^2)+ \int_{|z|^2}^{\infty}e^{-t}t^{k_1} \log tdt}{k_1!} \nonumber \\
& - \sum_{k_1=0}^{N-1}\frac{|z|^{2k_1}}{2k_1!}(k_1+1 - |z|^2) \Bigg \} + \mathcal{O}\big((\Gamma-2)^2\big)
\end{align}
(in the second last sum the term $\gamma(k_1+1, |z|^2)$ as presented in \cite{TF12} contains a typographical error and reads with $k_1$ in the argument instead of $k_1+1$), where $ \rho_{(1)}^{\rm d} (\vec{r};2)$ is given by (\ref{eq:C2}).

We seek the limiting form of the $\mathcal{O}(\Gamma-2)$ correction term as presented above under the edge scaling
\begin{equation}
  \label{eq:zy}
  z:=\sqrt{N}-y ,
\end{equation}
which effectively positions the neutralizing background of the plasma in the half plane $y>0$. For this task we hypothesize that in the limit $N \to \infty$, only the large $k_1, k_2$ portion of the sums in (\ref{eq:C1}) contribute, allowing us to use the asymptotic expansions
\begin{equation}
\label{eq:a1}
  \gamma(N-j+1; |z|^2)  \sim \frac{1}{2} \Gamma (N-j+1) \Big(1+{\rm erf}\big(\frac{j-2y\sqrt{N}}{\sqrt{2N}}  \big) \Big) 
\end{equation}
\begin{equation}
  \label{eq:a2}
  \Gamma (N-j+1; |z|^2)  \sim \frac{1}{2} \Gamma (N-j+1) \Big(1-{\rm erf}\big(\frac{j-2y\sqrt{N}}{\sqrt{2N}}  \big) \Big) 
\end{equation}
\begin{equation}
  \label{eq:a3}
  I(k_1, k_2)  \sim  \frac{k_1!k_2!}{2} {\rm erfc}\bigg(\frac{k_2-k_1}{\sqrt{2(k_1+k_2)}}\bigg). 
\end{equation}
The first two of these are standard results while the third was derived in \cite{TF12}. We remark that the asymptotic expansion of $J(k_1, k_2)$ follows by substituting (\ref{eq:a3})  in (\ref{eq:JI}), together with Stirling's formula. Thus we have
\begin{align}
  \label{eq:a4}
J(k_1, k_2) - k_1!k_2!\log N & \sim \bigg(\frac{1}{2} \frac{1}{k_1} + \log \frac{k_1}{N}  \bigg)I(k_1, k_2) + \bigg(\frac{1}{2} \frac{1}{k_2} + \log \frac{k_2}{N}  \bigg)I(k_2, k_1)\nonumber \\
& \quad + \frac{1}{\sqrt{2(k_1+k_2)}} \bigg(\frac{2}{\sqrt{\pi}} \bigg)k_1!k_2!e^{-(k_1-k_2)^2/(2(k_1+k_2))}.
\end{align}

With these preliminaries let us consider the scaled limit of the first double sum in (\ref{eq:C1})

\begin{lemma}\label{LM1}
  Under the assumption that the leading asymptotic portion of the sum comes from the large $k_1, k_2$ region, for large $N$ and with $z$ given by (\ref{eq:zy}) we have
  \begin{align}\label{eq:X1}
   & e^{-|z|^2} \sum_{k_2=0}^{N-1} \frac{1}{2k_2!} \sum_{k_1=0}^{N-1} \Big(\frac{1}{2} \frac{1}{k_1} + \log \frac{k_1}{N} \Big) I(k_1, k_2) \frac{1}{k_1!k_2!} \nonumber\\
&  \sim - \frac{1}{2} \frac{1}{\sqrt{2\pi}} \int_0^\infty dt_2 \, e^{-2(t_2-y)^2}(1+ {\rm{erf}} \, t_2) \nonumber \\
& \qquad + \frac{1}{\sqrt{2\pi}} \int_0^\infty dt_2 \, e^{-2(t_2-y)^2} \bigg(t_2^2 {\rm erfc} \, t_2 - \frac{t_2e^{-t_2^2}}{\sqrt \pi}  \bigg) \nonumber \\
& \qquad + \frac{e^{-|z|^2}}{2}\sum_{k_2=0}^{N-1} \frac{|z|^{2k_2}}{k_2!} \frac{(N-k_2)^2}{2N} .
  \end{align}
\end{lemma}

\noindent
Proof. \quad Consider the sum over $k_1$. After substituting (\ref{eq:a3}), breaking the sum up into the regions $k_1 \in [0, \ldots, k_2-1]$ and $k_1 \in [k_2, \ldots, N-1]$, writing
  \begin{equation*}
    \frac{1}{2}{\rm erfc}\bigg(\frac{k_2-k_1}{\sqrt{2(k_1+k_2)}} \bigg) = 1 - \frac{1}{2} {\rm erfc}\bigg(\frac{k_1-k_2}{\sqrt{2(k_1+k_2)}} \bigg)
  \end{equation*}
in the latter, and changing summation labels $k_1 \mapsto N- k_1, \ k_2 \mapsto N- k_2$ we see that
\begin{align*}
  & \sum_{k_1=0}^{N-1} \Big(\frac{1}{2} \frac{1}{k_1} + \log \frac{k_1}{N} \Big) I(k_1, k_2) \frac{1}{k_1!k_2!} \\
& \sim \frac{1}{2} \sum_{k_1=k_2}^{N} \Big(\frac{-k_1}{N} \Big) {\rm erfc} \Big(\frac{k_1-k_2}{2\sqrt{N}} \Big) - \frac{1}{2} \sum_{k_1=1}^{k_2-1} \Big(\frac{-k_1}{N} \Big) {\rm erfc} \Big(\frac{k_2-k_1}{2\sqrt{N}} \Big) \\
& \quad + \sum_{k_1=k_2}^{N-1} \Big(\frac{1}{2} \frac{1}{k_1} + \log \frac{k_1}{N} \Big) \\
& \sim - \frac{1}{2}\Big(1+{\rm erf}\frac{t_2}{2} \Big) + t_2^2 {\rm erfc} \, \frac{t_2}{2} - \frac{t_2e^{-t_2^2/4}}{2\sqrt N}-\frac{k_2^2}{2N}. 
\end{align*}
In the last line $t_2:= k_2/\sqrt N$, and this line is obtained from the line before by regarding the first two sums as Riemann sums, and by calculating the leading behaviour of the third sum.

Now performing the sum over $k_2$, using the asymptotic expression
$$
e^{-|z|^2} \frac{|z|^{2k_2}}{k_2!} \Big  |_{k_2 \mapsto N-k_2} \sim \frac{e^{-2y^2 + 2t_2y - t_2^2/2}}{\sqrt{2\pi N}}
$$
in the first two sums, which are again Riemann sums, and changing variables gives (\ref{eq:X1}). \hfill $\square$

\begin{lemma}\label{LM2}
  With the hypothesis of Lemma \ref{LM1}, 
  \begin{align*}
   & e^{-|z|^2} \sum_{k_2=0}^{N-1}\frac{1}{2k_2!}\sum_{k_1=0}^{N-1} \Big(\frac{1}{2} \frac{1}{k_2} + \log k_2 \Big)  \frac{I(k_1, k_2)}{k_1!k_2!} \\
& \sim  e^{-|z|^2} \Bigg( \sum_{k_2=0}^{N-1} \frac{|z|^{2k_2}}{4k_2!} +  \sum_{k_2=0}^{N-1} \frac{|z|^{2k_2}}{2k_2!}(k_2-N)+ \sum_{k_2=0}^{N-1} \frac{|z|^{2k_2}}{4k_2!} \frac{(k_2-N)^2}{N} \Bigg).
  \end{align*}
\end{lemma}

\noindent
Proof. \quad The derivation follows analogous reasoning to that of Lemma \ref{LM1}. 
\hfill $\square$

Substituting the final term on the RHS of (\ref{eq:a4}) in the first double sum of (\ref{eq:C1}) leads immediately to a Riemann sum and so its leading asymptotic behaviour is readily obtained. Combining this with the results of Lemmas 1 and 2, and taking into consideration too that the terms $k_2=k_1$ in the first term of (\ref{eq:C1}) are to be excluded, gives the following form of the leading behaviour. 

\begin{lemma}
  With the hypothesis of Lemma \ref{LM1}, 
  \begin{align*}
   & e^{-|z|^2} \sum_{k_1=0}^{N-1} \sum_{\substack{k_2=0 \\ k_2 \ne k_1} }^{N-1} \frac{|z|^{2k_2}}{2k_1!(k_2!)^2}J(k_1, k_2)\\
& \sim -\frac{e^{-|z|^2}}{2} \sum_{k=1}^{N-1}\Big(\frac{1}{ \sqrt{\pi k}} +\log k  \Big) \frac{|z|^{2k}}{k!}\\
& \quad -\frac{1}{4}  e^{-|z|^2} \sum_{k_1=0}^{N-1} \frac{|z|^{2k_1}}{k_1!} + (2 \sqrt{N}y + y^2) \frac{e^{-|z|^2}}{2}\sum_{k_1=0}^{N-1}  \frac{|z|^{2k_1}}{k_1!} \\
& \quad + \frac{1}{\sqrt{2\pi}} \int_{0}^\infty dt_2 \, e^{-2(t_2-y)^2}t_2 \Big(\frac{e^{-t_2^2}}{\sqrt \pi} - t_2 {\rm erfc}\, t_2 \Big) \\
& \quad + N \log N e^{-|z|^2}  \sum_{k_2=0}^{N-1}  \frac{|z|^{2k_2}}{2k_2!}.
  \end{align*}
\end{lemma}

The scaled large $N$ form of the second and third terms in (\ref{eq:C1}) follows upon substituting $(\ref{eq:a1})-(\ref{eq:a3})$ as appropriate, and observing that Riemann sums result.

\begin{lemma}
  Under the hypothesis of Lemma \ref{LM1}
  \begin{align*}
    & \sum_{k_1=0}^{N-1} \sum_{k_2=k_1+1}^{N-1} \frac{I(k_1,k_2)}{k_1!k_2! (k_2-k_1)} \Big(\frac{|z|^{2k_2}}{k_2!} +  \frac{|z|^{2k_1}}{k_1!} \Big)\\
&\quad  - \sum_{k_1=0}^{N-1} \sum_{k_2=k_1+1}^{N-1} \frac{|z|^{2k_1}\gamma (k_2+1, |z|^2) + |z|^{2k_2} \Gamma(k_1+1, |z|^2)}{(k_2-k_1)k_1!k_2!} \\
& \sim - \frac{1}{\sqrt{2\pi}} \int_{0}^\infty dt_1 \int_{0}^{t_1}dt_2 \frac{1}{t_1-t_2} \bigg( e^{-2 (t_1 - y)^2} \Big({\rm erf} (t_1-t_2) + {\rm erf}\big(\sqrt{2} (t_2 - y) \big) \Big)\\
& \quad  + e^{-2 (t_2 - y)^2} \Big({\rm erf} (t_1-t_2) - {\rm erf}\big(\sqrt{2} (t_1 - y) \big)  \Big)  \bigg).
  \end{align*}
\end{lemma}

Regarding the final double sum in (\ref{eq:C1}) we first observe
\begin{align}
\label{eq:im}
  - &\sum_{k_2=0}^{N-1}  \frac{|z|^{2k_2}}{2k_2!} \sum_{k_1=0}^{N-1} \frac{\gamma (k_2+1, |z|^2)\log |z|^2 + \int_{|z^2|}^\infty e^{-t}t^{k_1} \log t dt}{k_1!}
  \nonumber \\
  & = - {N \over 2} \log |z|^2 \sum_{k_2=0}^{N-1} {|z|^{2k_2} \over k_2!} 
   - \sum_{k_2=0}^{N-1} {|z|^{2k_2} \over 2 k_2 !}
  \sum_{k_1 = 0}^{N-1} 
  {\int_{|z|^2}^\infty e^{-t} t^{k_1} \log {t \over |z|^2} \, dt \over k_1!}.
\end{align}
The saddle point method can be used to obtain the asymptotic form of the sum over $k_1$ on the RHS. Doing this shows that a Riemann sum results. Furthermore, the resulting integral can be exactly evaluated. Taking into consideration too that the term $k_1 = k_2$ is excluded in the final double sum in (\ref{eq:C1}) we obtain the following result.

\begin{lemma}\label{LM5}
  Under the hypothesis of Lemma \ref{LM1}
  \begin{align*}
    & \sum_{k_1=0}^{N-1} \sum_{\substack{k_2=k_1+1 \\ k_1 \ne k_2} }^{N-1} \frac{|z|^{2k_1}\gamma (k_2+1, |z|^2)+ |z|^{2k_2}\Gamma(k_1+1, |z|^2)}{(k_2-k_1)k_1!k_2!}\\
& \sim (\log N) e^{-|z|^2} \sum_{k_1=0}^{N-1} \frac{|z|^{2k_1}}{2k_1!}\\
& \quad -\frac{1}{4} \big(1+ {\rm erf}(\sqrt{2}y) \big) \Big \{ \frac{ye^{-2y^2}}{\sqrt{2\pi}} + \big(\frac{1}{4} +y^2  \big) \big(1+ {\rm erf}(\sqrt{2}y) \big) \Big \}.
  \end{align*}
  \end{lemma}

Substituting the results of Lemmas \ref{LM1} to \ref{LM5} in (\ref{eq:C1}) gives the sought scaled limit of the $\mathcal{O}\big((\Gamma- 2)\big)$ correction to the edge scaled density $\rho _{(1)}^{\rm edge} (y; \Gamma)$.

\begin{prop}
  We have
  \begin{equation}\label{eq:7.0}
    \rho _{(1)}^{\rm edge} (y; \Gamma) = \rho _{(1)}^{\rm edge} (y;2)  - {(\Gamma -2) \over \pi} A(y) + \mathcal{O}\big((\Gamma- 2)^2\big),
  \end{equation}
where
  \begin{equation}\label{eq:7.1}
    \rho _{(1)}^{\rm edge} (y;2) = \frac{1}{2\pi} {\rm erfc} (-\sqrt{2}y)
  \end{equation}
and $A(y) = A_1(y) + A_2(y) + A_3(y) + A_4(y)$ with 
\begin{align*}
  A_1(y) = & - \frac{1}{2\sqrt{2\pi}} \int_0^\infty e^{-2(t-y)^2} {\rm erfc}\,t \, dt \\
  A_2(y) = & \frac{1}{\sqrt{2\pi}} \int_0^\infty e^{-2(t-y)^2} t \big(\frac{e^{-t^2}}{\sqrt{\pi}} - t {\rm erfc}\,t \big) \,dt\\
  A_3(y) = & \frac{1}{4}\big(1+ {\rm erf}(\sqrt{2}y)\big) \Big ( - \frac{ye^{-2y^2}}{\sqrt{2\pi}} + \big(\frac{1}{4} +y^2  \big)\big(1 - {\rm erf}(\sqrt{2}y) \big)\Big)\\
  A_4(y) = & -\frac{1}{\sqrt{2\pi}}\int_0^\infty dt_1 \int_0^{t_1}dt_2 \frac{1}{t_1-t_2} \\ 
& \times \bigg(e^{-2(t_2-y)^2} \Big({\rm erf} (t_1-t_2) + {\rm erf} \big(\sqrt{2}(t_2-y) \big) \Big)\\
& + e^{-2(t_2-y)^2} \Big({\rm erf} (t_1-t_2) - {\rm erf} \big(\sqrt{2}(t_1-y) \big)  \Big) \bigg).
\end{align*}
\end{prop}

\subsection{Cylinder geometry}\label{AA4}

The leading order correction to the density at \(\Gamma = 2\) in the soft cylinder geometry for finite \(N\) for a droplet with mean density \(N/(LW) = \rho_{b} \) is 
\begin{align}\label{KA1}
\rho_{(1)}^{c}(y; \Gamma) & = \rho_{(1)}^{c}(y; 2)\nonumber \\
&  - (\Gamma-2) \Bigg\{    -\frac{1}{2}  \partial_{y} \left( y \rho_{(1)}^{c}(y; 2)\right)  -\frac{1}{16\pi \rho_b}\partial_{y}^{2}\rho_{(1)}^{c}(y;2) \nonumber \\
&\qquad \qquad \quad+\frac{ \sqrt{\pi}}{ W^{2}} \sum_{0 \le a\ne b < N} e^{- 2\pi \rho_b \left(y - 
k_{a}\right)^{2}}
\left[\sqrt{2} F\left(\sqrt{\pi \rho_b}(k_{a} - k_{b})\right)  - F\left(\sqrt{2\pi \rho_b}( y - k_{b})\right) \right] \nonumber \\
&\qquad \qquad \quad +  \frac{ 1}{W^{2}\sqrt{2 \rho_b} } \sum_{0\le a\ne b < N}\frac{e^{- 2\pi \rho_b \left(y- k_{a}\right)^{2}}}{k_a-k_b}
\left[{\rm erf}\left(\sqrt{2\pi \rho_b}(y
- k_{b})\right)- {\rm
erf}\left(\sqrt{\pi \rho_b}(k_{a} - k_{b})\right)\right]\Bigg\} \nonumber \\
&\qquad \qquad \quad + \mathcal{O}\left( (\Gamma-2)^{2}\right).
\end{align}
Here \(k_{n} \equiv \frac{n}{W \rho_b} \), \(F(x) \equiv x\, \left(1 +  {\rm erf}(x)\right) + e^{- x^{2}}/\sqrt{\pi}\), and the particle density at \(\Gamma = 2\) is given by (\ref{F12}).

The \(y\) coordinate here is chosen such that one edge of the droplet is at \(y = 0\) for all \(N\), making it a natural parameterization for studying the limiting edge density. Indeed, the limiting edge density for the soft disk (\ref{eq:7.0}) is recovered in the limit \(N, W, L \to \infty\) for fixed \(y\) and \(L/W = \mathcal{O}(1)\). The droplet for \(\Gamma = 2\) occupies the region \( 0 \le x \le W\) and \(0 \le y \le L\), and the leading correction is localized to distances on the order of the magnetic length \(l_{B} = \left( 2\pi \rho_{b}\right)^{-1/2}\) from each edge when \(W \gg  l_{B}\). We remark that in the thin cylinder limit \(W \sim  l_{B}\), the correction develops oscillatory features which extend into the bulk. 

The derivation of (\ref{KA1}) closely mirrors that of the leading order correction for the disk geometry presented in Ref.~\cite{TF12}, with only minor changes reviewed below. Writing the correction as \(\rho_{(1)}^{c}(y; \Gamma) = \rho_{(1)}^{c}(y; 2) - \frac{(\Gamma - 2)}{2} \langle \hat{\rho}(\vec{r}) U \rangle^{T}\) where \(U\) is the total potential energy of the plasma, and the truncated average is taken with the Boltzmann factor at \(\Gamma = 2\), we get
\begin{align*}
\langle \hat{\rho}(\vec{r}) U \rangle^{T} &= 2\pi \rho_b \,y^{2} \rho_{(1)}^{c}(\vec{r}) +2\pi \rho_b  \int_{\Omega} d^{2} \vec{r}_2   \left[ \rho_{(2)}^{c}(\vec{r},\vec{r}_2) - \rho_{(1)}^{c}(\vec{r})\rho_{(1)}^{c}(\vec{r}_2)\right] y_{2}^{2}  + 2\int_{\Omega} d^{2} \vec{r}_{2}  \rho_{(2)}^{c}(\vec{r},\vec{r}_2)v(\vec{r},\vec{r}_2)\\
& + \int \int_{\Omega \times \Omega} d^{2} \vec{r}_{2}  d^{2} \vec{r}_{3}  \left[ \rho_{(3)}^{c}(\vec{r},\vec{r}_2,\vec{r}_3) - \rho_{(1)}^{c}(\vec{r}) \rho_{(2)}^{c}(\vec{r}_2,\vec{r}_3)\right] v(\vec{r}_2,\vec{r}_3), 
\end{align*}
\(v(z, z') = -\log | e^{2\pi i \bar{z}/W} - e^{2 \pi i \bar{z}' / W} |\), and the domain of integration \(\Omega := \{ (x, y)| x \in [0, W), y \in \mathbb R\} \). The form of the ``potential" and neutralizing background potential is chosen to emphasize the analogy with the disk geometry. To relate this back to the 2D Coulomb plasma on a cylinder, note that replacing \(v(z,z') \to \Phi(z, z')\) in the expression above, and translating coordinates \(y \to y - L(N-1)/2N\), will leave the left hand side \(\langle \hat{\rho} U\rangle^{T}\) unchanged.

At \(\Gamma  =2\) in the soft cylinder geometry, the correlation functions needed to calculate the correction have the structure \cite{CFS83}
\begin{align}\label{CA1}
\rho_{\ell}^{c}(\vec{r}_1, \vec{r}_2, ..., \vec{r}_{\ell}) = \rho_b^{\ell} \det\left( K(\vec{r}_{i}, \vec{r}_{j})\right)_{1 \le i,j\le \ell},
\end{align}
where
\begin{align}\label{CA2}
K(\vec{r}_{1}, \vec{r}_{2}) &= \frac{1}{W} \sqrt{\frac{2}{ \rho_{b}}}  \sum_{n = 0}^{N-1} e^{2 \pi i n(x_1 - x_2)/W} e^{- \pi \rho_b (y_1 - k_n)^{2} - \pi \rho_b (y_2- k_n)^{2}},
\end{align}
and \(k_n\) is defined above. Explicit evaluation of the correction is further facilitated by expanding the ``potential" in a Fourier series in the periodic direction
\begin{align}\label{CA3}
v(z, z') = - \frac{2\pi}{W} \max(y, y') + \sum_{m = 1}^{\infty} \frac{1}{m} \cos\left( \frac{2\pi m }{W} (x - x')\right) e^{ - 2\pi m |y - y'|/W}.
\end{align}

After some lengthy calculations, analogous to those detailed above in the soft disk case and therefore omitted, the same limiting edge density  (\ref{eq:7.0}) as found for the soft disk is reclaimed.

\section{Large deviation and asymptotic edge density outside the droplet for general $\Gamma$}\label{SK}
\subsection{Introductory Remarks}

By definition a one-component plasma system consists of a smeared out, charge neutralizing background, and $N$ mobile charges. In the large $N$ limit the leading order density of mobile charges must coincide with the density of the background; if not the charge imbalance would create an electric field, and the system would not be in equilibrium.

We are interested in the situation that the mobile particles are free to move throughout the plane (soft disk) or cylinder, and furthermore that the potential they experience is the analytic continuation of that inside of the neutralizing background. Furthermore, scaled variables are to be used so that the leading support of the background is independent of $N$. In this setting for one-component log-gas systems on the line, Gaussian fluctuation formulas for linear statistics valid for general coupling have recently been used to calculate the leading (exponentially small in $N$) density outside of the neutralizing background \cite{Fo11a,Fo11b}.
 We seek to do the same for the two-dimensional one-component plasma, in scaled soft disk or cylinder geometry. 

In the scaled soft disk, with the support of the leading density the unit disk, and $a(\vec{r})$ smooth on this domain, the appropriate Gaussian fluctuation formula reads \cite{Fo99}  
\begin{equation}
  \label{eq:A1}
  \Big \langle \prod_{l =1}^{N} e^{ika(\vec{r}_l)} \Big  \rangle \mathop{\sim}_{N \to \infty} e^{ik \mu _N} e^{- k^2 \sigma^2 /2}
\end{equation}
where, with $\Omega$ the unit disk
\begin{equation}
  \label{eq:A2}
  \mu_N = \int_\Omega a(\vec{r})\rho_{(1)}(\vec{r}) \, d\vec{r},
\end{equation}
\begin{equation}
  \label{eq:A3}
  \sigma^2 = \sigma_{\rm bulk}^2 + \sigma_{\rm surface}^2,
\end{equation}
with 
\begin{equation}
  \label{eq:A4}
  \sigma_{\rm bulk}^2 = \frac{1}{2\pi \Gamma} \int_\Omega \Big(\big ( \frac{\partial a}{\partial x}  \big )^2 + \big ( \frac{\partial a}{\partial y}  \big )^2 \Big)dx dy
\end{equation}
and
\begin{equation}
  \label{eq:A5}
  \sigma_{\rm surface}^2 =  \frac{2}{\Gamma} \sum_{n=1}^{\infty} na_na_{-n}, \qquad a(\vec{r}) |_{r=1} = \sum_{n=-\infty}^{\infty} a_ne^{in\theta}.
\end{equation}
Rigorous proofs of (\ref{eq:A1}) in the case $\Gamma = 2$ have been given in \cite{RV07s,AHM08}.

Consideration of the derivation of (\ref{eq:A1}) for the scaled soft disk geometry given in \cite{Fo99} implies that for the scaled cylinder, with the leading support of the density confined to say the unit square, (\ref{eq:A1}) again holds true. Of course in (\ref{eq:A2}) and (\ref{eq:A4}), $\Omega$ is now the unit square on the cylinder, and in (\ref{eq:A2}) $\rho_{(1)}(\vec r)$ is the corresponding particle density. Furthermore, the boundary of $\Omega$ now consists of two components: $y = 0$ and $y=1$, so (\ref{eq:A3}) should be modified to read
\begin{equation}
  \label{eq:A6}
  \sigma^2 =  \sigma_{\rm bulk}^2 + \sigma_{{\rm surface}, 0}^2 + \sigma_{{\rm surface}, 1}^2 
\end{equation}
with 
\begin{equation}
  \label{eq:A7}
  \sigma_{{\rm surface}, j}^2 = \frac{2}{\Gamma} \sum_{n=1}^{\infty} na_n^{(j)}a_{-n}^{(j)}, \qquad a(\vec{r}) |_{y=j} = \sum_{n=-\infty}^{\infty} a_n^{(j)}e^{2\pi in x}.
\end{equation}

\subsection{Exact asymptotics for $\Gamma = 2$ and $\Gamma = 2+ \varepsilon$  $(\varepsilon \ll 1)$}\label{SE}
First we compute the large deviation form of the density in disk geometry for $\Gamma = 2$, or equivalently the asymptotic large $N$ form of the density outside the leading support.

\begin{lemma}
  In disk geometry for $\Gamma = 2$ we have, for $r>1 $, 
  \begin{equation}
    \label{eq:Ax}
    \rho_{(1)}^{N, d}(\sqrt{N}r) \sim \frac{e^{-N(r^2-1)} e^{2N \log r}}{\pi (2\pi N)^{1/2}(r^2-1)}.
  \end{equation}
\end{lemma}

 \noindent
  Proof. \quad From the definition, simple manipulation and use of integration by parts
show that for $z\gg a\gg 1$, 
\begin{equation*}
  \Gamma(a+1;z) \sim \frac{e^{-z}z^{a+1}}{z-a}.
\end{equation*}
Using this and Stirling's formula in (\ref{eq:C2}) gives (\ref{eq:Ax}). \hfill $\square$

We next present the analogous formula in the case of cylinder geometry.

\begin{lemma}
  In cylinder geometry for $\Gamma =2$, we have for $y < 0$ 
  \begin{equation}
    \label{eq:Ay}
    \rho_{(1)}^{N, c}(\sqrt{N}y) \sim \sqrt{\frac{2}{N}} e^{-2\pi Ny^2} \Big (
    \frac{e^{2\pi y}}{1 - e^{4 \pi y}} - {\pi \over 2 N} \frac{d^2}{dy^2} \frac{e^{2\pi y}}{1
    - e^{4 \pi y}} \Big ).
  \end{equation}
\end{lemma}

\noindent
Proof. \quad A minor rewrite of (\ref{F12}) in the case $\rho_b=1$, $W = \sqrt{N}$ shows 
\begin{equation*}
  \rho_{(1)}^{N, c}(\sqrt{N}y) = \sqrt{\frac{2}{N}} e^{-2\pi Ny^2} \sum_{a=0}^{N-1} e^{4 \pi y (a+1/2)}e^{-2 \pi (a+1/2)^2/N}.
\end{equation*}
Expanding the final exponential in powers of $1/N$ gives, upon recalling $y<0$, 
\begin{equation*}
 \rho_{(1)}^{N, c}(\sqrt{N}y) \sim \sqrt{\frac{2}{N}} e^{-2\pi Ny^2} \sum_{a=0}^{N-1} e^{4 \pi y (a+1/2)} \Big ( 1 - {2 \pi (a + 1/2)^2 \over N} + O(1/N^2) \Big ).
\end{equation*}
Extending the upper terminal of the summation to infinity gives (\ref{eq:Ay}).  \hfill $\square$

\medskip
Let us denote the RHS of (\ref{eq:Ax}) by $\tilde{\rho}_{(1)}^{N,d}(\sqrt{N}r) $. We see that
\begin{equation}
  \label{eq:U1}
  \lim_{N\to \infty} \tilde{\rho}_{(1)}^{N,d}(\sqrt{N}r)|_{r=1-y/\sqrt{N}} = \frac{e^{-2y^2}}{(2\pi)^{3/2}|y|}.
\end{equation}
Using analogous notation, it follows from (\ref{eq:Ay}) that
\begin{equation}
  \label{eq:U2}
  \lim_{N\to \infty} \frac{1}{\pi} \tilde{\rho}_{(1)}^{N,c}(\sqrt{N}y) \Big |_{y \mapsto y/\sqrt{N}} = \frac{e^{-2y^2}}{(2\pi)^{3/2}|y|}
\end{equation}
(the factor of $\frac{1}{\pi}$ on the LHS of (\ref{eq:U2}) accounts for the change in the measure $ydy$) thus reproducing the same scaled form. We see from (\ref{eq:7.1}) that this scaled form is precisely the $y \to - \infty$ asymptotic form of (\ref{eq:7.1}),
\begin{equation}
  \label{eq:U3}
  \rho_{(1)}^{\rm edge} (y; 2) \mathop{\sim }_{y \to -\infty} \frac{e^{-2y^2}}{(2\pi)^{3/2}|y|}.
\end{equation}

In the next subsection, the Gaussian fluctuation formula (\ref{eq:A1}) will be used to compute $\tilde{\rho}_{(1)}^{(N), c}(\sqrt{N}y)$ and $\tilde{\rho}_{(1)}^{N, d}(\sqrt{N}r)$ for general $\Gamma > 0$. By scaling as in (\ref{eq:U2}) and (\ref{eq:U1}) respectively we find that the same scaled form results, and this scaled form is expected to be $y \to - \infty$ asymptotic form of $\rho_{(1)}^{\rm edge}(y; \Gamma)$. A test on this latter prediction is to expand it about $\Gamma = 2$ to first order in $\varepsilon := \Gamma - 2$, and compare it with the exact expansion of the $y \to - \infty$  asymptotic form of $\rho_{(1)}^{\rm edge}(y; \Gamma)$ as computed from (\ref{eq:7.0}). 

\begin{lemma}
  With $\varepsilon := \Gamma - 2$ and $\rho_{(1)}^{\rm edge}(y; \Gamma) - \rho_{(1)}^{\rm edge}(y; 2) := -\frac{\e}{\pi} A(y) + \mathcal{O}(\e^2)$, $A(y)$ as in (\ref{eq:7.0}), we have
  \begin{equation}
    \label{eq:U4}
    A(y) \mathop{\sim}_{y \to -\8} |y| \frac{e^{-2y^2}}{(2\pi)^{1/2}}.
  \end{equation}
\end{lemma}

\noindent
Proof. \quad A detailed consideration of the $y \to -\8$ asymptotic form of $ A(y)$ is given in Appendix \ref{AA5}. To leading order one has that $ A(y) \mathop{\sim}_{y \to -\8} A_3(y)$. But
\begin{equation*}
  A_3(y) \mathop{\sim}_{y \to -\8} \frac{y^2}{2} {\rm erfc}(\sqrt{2}|y|),
\end{equation*}
thus implying (\ref{eq:U4}) \hfill $\square$.\\
\\
Finally, to complete the discussion of exact asymptotics, we present the asymptotes inside the droplet. First, we need the following lemma.

\begin{lemma}\label{lem:ode}
The antisymmetric part of \(A(y)\) as in (\ref{eq:7.0}), denoted by \(A_{a}(y) = \frac{1}{2}(A(y) - A(-y))\), obeys the ordinary differential equation

\begin{equation}\label{eq:ode1}
A_{a}''(y) + 4 y A_{a}'(y) =  \left(2y^{2} - \frac{1}{2}\right)\left( {\rm erf}(\sqrt{2/3}y)-{\rm erf}(\sqrt{2}y)\right)  + \frac{ \sqrt{6}}{ \sqrt{\pi}} y e^{-2 y^{2}/3}.
\end{equation}
\end{lemma}

Proof. This follows most readily by noting that the LHS is equivalent to \(e^{- 2y^{2}} \partial_{y} \left( e^{2y^{2}} \partial_{y} A_{a}(y)\right)\), and applying this operation in the sequence implied to \(A_{a}\). Details of this computation are presented in Appendix \ref{AA6}.

\begin{lemma}
  With $\varepsilon := \Gamma - 2$ and $\rho_{(1)}^{\rm edge}(y; \Gamma) - \rho_{(1)}^{\rm edge}(y; 2) := -\frac{\e}{\pi} A(y) + \mathcal{O}(\e^2)$, $A(y)$ as in (\ref{eq:7.0}), we have that inside the droplet, 
  \begin{equation}
    \label{eq:Ain}
    A(y) \mathop{\sim}_{y \to \8} -\frac{1}{2 \sqrt{\pi}} \left( \frac{3}{2}\right)^{5/2} \frac{e^{- 2y^{2}/3}}{y^{3}}.
  \end{equation}
\end{lemma}

\noindent
Proof. Expanding the RHS of Eq. (\ref{eq:ode1}) for large \(y\) gives
\begin{align*}
A_{a}''(y) + 4 y A_{a}'(y) \mathop{\sim}_{y \to \8}  \frac{2 \sqrt{3}}{\sqrt{2\pi}} \frac{e^{- 2y^{2}/3}}{y},
\end{align*}
which admits the asymptotic solution
\begin{align*}
A_{a}(y)  = - \frac{1}{4 \sqrt{\pi}}\left(\frac{3}{2}\right)^{5/2} \left(1+ \mathcal{O}\left( y^{-1}\right)\right) \frac{e^{- 2 y^{2}/3}}{y^{3}}.
\end{align*}

Since the density decays like \(e^{- 2y^{2}}\) outside the droplet, the dominant contribution to the large \(y\) behavior of \(A_{a}(y)\) must come from the interior asymptote, implying \(A(y) \mathop{\sim}_{y\to \8} 2 A_{a}(y)\) and thus (\ref{eq:Ain}).  \hfill $\square$

\subsection{Gaussian fluctuation formula predictions}
We will consider first the soft disk 2dOCP. To specify the particle density, we require the configuration integral (\ref{Q1}).
In terms of this notation, for the system with background density $\rho_b = 1/\pi$ and $N+1$ particles, we have 
\begin{align}
\label{eq:Av}
  \rho_{(1)}^{N+1, {\rm d}}(\vec{r}) &  = (N+1) \frac{e^{-(\Gamma/2)r^2}}{Q_{N+1,\Gamma}^d(1/\pi)} \int_{{\mathbb R}^2}d\vec{r_1} \cdots \int_{{\mathbb R}^2}d\vec{r_N} \, e^{-(\Gamma /2)\sum_{j=1}^N \vec{r_j}^2} \nonumber \\
& \quad \times \prod_{l = 1}^N|\vec{r}- \vec{r_l}|^\Gamma \prod_{1 \leq j < k \leq N} |\vec{r_k}- \vec{r_j}|^\Gamma \nonumber \\
& = (N+1) e^{-(\Gamma /2)r^2} \frac{Q_{N,\Gamma}^d(1/\pi)}{Q_{N+1,\Gamma}^d(1/\pi)}  \Big \langle \prod_{l=1}^{N}|\vec{r}- \vec{r_l}|^\Gamma  \Big \rangle_{\widehat{\rm IQ}_{N, \Gamma}^d (1/\pi)},
\end{align}
where $\widehat{\rm IQ}_{N, \Gamma}^d (\rho_b)$ refers to the PDF corresponding to the integrand of $Q_{N, \Gamma}^d (\rho_b)$. Furthermore, changing variables $\vec{r_l} \mapsto \sqrt{N} \vec{r_l}$ in (\ref{eq:Av}) shows 
\begin{equation}\label{eq:Av1}
  \rho_{(1)}^{N+1, {\rm d}}(\sqrt{N+1}\vec{r}) = (N+1)N^{\Gamma N/2} e^{-(N+1)\Gamma r^2/2} \frac{Q_{N,\Gamma}^{\rm d}(1/\pi)}{Q_{N+1,\Gamma}^{\rm d}(1/\pi)}  \Big \langle \prod_{l=1}^{N}\Big |\sqrt{\frac{N+1}{N}}\vec{r}- \vec{r_l} \Big |^\Gamma  \Big \rangle_{\widehat{\rm IQ}_{N, \Gamma}^{\rm d} (N/\pi)}.
\end{equation}

We recognise the average in (\ref{eq:Av}) as an example of the LHS of (\ref{eq:A1}) with 
\begin{equation}
  \label{eq:ka}
  k = -i\Gamma, \qquad a(\vec{r_l}) = \log \Big |\sqrt{\frac{N+1}{N}}\vec{r}- \vec{r_l} \Big |.
\end{equation}
Our task then is to compute $\mu_N$ and $\sigma^2$ appearing in the RHS of (\ref{eq:A1}), as specified by (\ref{eq:A2}) -- (\ref{eq:A5}).

\begin{lemma}
  For the soft disk with $\rho_b = N/\pi$, $a(\vec{r_l})$ as in (\ref{eq:ka}), and with $r>1$ we have
  \begin{equation}
    \label{eq:B1}
    \mu_N = N\log r + \frac{1}{2} + o(1)
  \end{equation}
  \begin{equation}
    \label{eq:B2}
    \sigma_{\rm bulk}^2 = \sigma_{\rm surface}^2 = - \frac{1}{2} \log \big(1 - \frac{1}{r^2} \big).
  \end{equation}
\end{lemma}

\noindent
Proof. \quad Let $\rho_{(1)}^{N,g}(r)$ denote the global density in the soft disk plasma system with $\rho_{b}=N/\pi$. {{} Generally the global density for log-potential system refers to the
density that results from scaling the variables so that the leading order support is a finite
domain.} We know from \cite{ZW06},
\cite[below (5.16) and (5.17)]{TF12} that this has the large $N$ form
\begin{equation}
  \label{eq:B3}
  \rho_{(1)}^{N,g}(r) = \frac{N}{\pi} \chi_{0<r<1}+ \frac{1}{2\pi \Gamma} \big(1- \frac{\Gamma}{4} \big)\frac{1}{r} \delta ' (r-1) + o(1),
\end{equation}
where $\chi_J = 1$ for $J$ true, $\chi_J = 0$ otherwise. Substituting in (\ref{eq:A2}), (\ref{eq:B1}) results after an elementary calculation.

Choosing, without loss of generality, $\vec r = (\tilde{r}, 0)$, $\tilde{r} := \sqrt{\frac{N+1}{N}}r$ and $\vec {r_l} = (x, y)$ in the definition (\ref{eq:ka}) of $a(\vec {r_l})$ and substituting in (\ref{eq:A4}) shows after some simple computation and the introduction of polar coordinates, that
\begin{align}
\label{eq:s1}
  \sigma_{\rm bulk}^2 & = \frac{1}{2\pi \Gamma} \int_0^1 dR\  R\int_0^{2\pi}d\theta \frac{1}{(\tilde{r} - Re^{i\theta})(\tilde{r} - Re^{-i\theta})} \nonumber \\
& = - \frac{1}{2 \Gamma} \log \big(1 - \frac{1}{\tilde r^2} \big). 
\end{align}

In relation to the computation of $\sigma_{\rm surface}^2$, similarly without loss of generality we can write
\begin{equation*}
  a(\vec {r_l}) = \log \tilde{r} + \log \Big |1 - \frac{\vec {r_l}}{\tilde r} \Big |
\end{equation*}
thus telling us 
\begin{equation*}
  a_n = a _{-n} = \frac{1}{2n\tilde r^n}, \qquad (n \neq 0).
\end{equation*}
Consequently
\begin{equation}
  \label{eq:s2}
  \sigma_{\rm surface}^2 = \frac{1}{2 \Gamma} \sum_{n=1}^\8 \frac{1}{n\tilde r^{2n}} = - \frac{1}{2 \Gamma} \log \big(1 - \frac{1}{\tilde r^2} \big).
\end{equation}
Adding together (\ref{eq:s1}) and (\ref{eq:s2}) gives
\begin{equation}
  \label{eq:s3}
  \sigma^2 = - \frac{1}{\Gamma} \log \big(1 - \frac{1}{\tilde r^2} \big).
\end{equation}
 \hfill $\square$

\medskip
Now substituting the result of Lemma 9 in the RHS of (\ref{eq:A1}) with $k$ as in (\ref{eq:ka}) we see that
\begin{equation}
  \label{eq:s4}
\Big \langle \prod_{l=1}^N \big |\sqrt{\frac{N+1}{N}} \vec{r} -  \vec{r_l} \big |^\Gamma   \Big \rangle _{\widehat{\rm IQ}_{N, \Gamma}^d (N/\pi)} = \exp \big( N\Gamma \log r+ \frac{\Gamma}{2} -  \frac{\Gamma}{2} \log (1 - \frac{1}{r^2})+ o(1) \big).
\end{equation}

With regards to the large $N$ form of the ratio of partition functions in (\ref{eq:Av1}) we note from the explicit form of the Boltzmann factor (\ref{eq:d3}) that the dimensionless free energy is given by
\begin{equation}
  \label{eq:17.1}
  \beta F_N (\Gamma, \rho_b)|_{\rho_b=1/\pi} = - \log \frac{1}{N!} Q_{N, \Gamma}^d + \frac{\Gamma}{2} \big(\frac{N^2}{2} \log N - \frac{3N^2}{4}  \big).
\end{equation}
The free energy for the $2d {\rm OCP}$ is extensive \cite{SM76}  and thus for large $N$
\begin{equation}
  \label{eq:17.2}
  \beta F_{N+1} (\Gamma, \rho_b) -  \beta F_{N} (\Gamma, \rho_b) = \beta f (\Gamma, \rho_b) + o(1),
\end{equation}
(recall (\ref{k1})). Substituting (\ref{eq:17.1}) in (\ref{eq:17.2}) shows
\begin{equation}
  \label{eq:17.3}
  \log \frac{(N+1)Q_{N, \Gamma}^{\rm d} (1/\pi)}{Q_{N+1, \Gamma}^d (1/\pi)} = -  \frac{\Gamma}{2} \big( (N+\frac{1}{2}) \log N -N  \big) +  \beta f (\Gamma, 1/\pi) + o(1).
\end{equation}

Substituting (\ref{eq:s4}) and (\ref{eq:17.3}) in (\ref{eq:Av}1) gives our sought large deviation formula.

\begin{prop}
  For the soft disk $2d{\rm OCP}$ with $\rho_b = 1/\pi$ and corresponding dimensionless free energy per particle $\beta f (\Gamma, \rho_b)$ we have for $r>1$
  \begin{equation}
    \label{eq:17.4}
    \rho_{(1)}^{N,{\rm d}}(\sqrt{N}r) = \frac{e^{\beta f (\Gamma, 1/\pi)}}{N^{\Gamma /4}}e^{-(N\Gamma /2)(r^2-1)}  \exp \big(N\Gamma \log r - \frac{\Gamma}{2} \log (r^2-1) + o(1) \big).
  \end{equation}
\end{prop}

For $\Gamma = 2$ we can check (\ref{eq:17.4}) against the exact result (\ref{eq:Ax}). Thus for $\Gamma = 2$ we read off from
(\ref{F2a}) that
{{} $\beta f (2, 1/\pi) = \frac{1}{2} \log(1/2\pi^3)$}. Substituting this in (\ref{eq:17.4}) with $\Gamma = 2$ indeed reclaims (\ref{eq:Ax}). 

We now turn our attention to deriving the analogue of Proposition 2 for cylinder geometry. With $\rho_b = N/LW$ the appropriate configuration integral is
(\ref{Q2}),
and analogous to (\ref{eq:Av1}), in a system of $(N+1)$ particles the corresponding particle density can be written
\begin{align}
\label{eq:Bv1}
  \rho_{(1)}^{(N+1),{\rm c}}(y) =  (N+1) \frac{Q_{N, \Gamma}^c(L,W)}{Q_{N+1, \Gamma}^{\rm c}(L,W)}e^{-\Gamma \pi \rho_b (y- W/2)^2}
 \Big \langle \prod_{j=1}^N \big |2\sin \frac{\pi}{L} \big (x_j + i(y-y_j) \big) \big |^\Gamma \Big \rangle_{\widehat{IQ}_{N, \Gamma}^{\rm c} (L,W)}.
\end{align}
And if we further specialize to the case that $\rho_b = 1$, $L = W =\sqrt{N+1}$ (\ref{eq:Bv1}) can be rewritten, upon simple changes of variables
\begin{align}
  \label{eq:Bv2}
 \rho_{(1)}^{(N+1),{\rm c}}(\sqrt{N+1}\,y) = & N(1+ \frac{1}{N})^N \frac{Q_{N, \Gamma}^{\rm c}(\sqrt{N}, \sqrt{N})}{Q_{N+1, \Gamma}^{\rm c}(\sqrt{N+1},\sqrt{N+1})}e^{-\Gamma \pi (N+1) (y- 1/2)^2} \nonumber \\
& \times  \Big \langle \prod_{l=1}^N e^{-\Gamma \pi (y_l - 1/2)^2} \big |2\sin \big( \pi x_l + \pi i(y-y_l) \big) \big |^\Gamma \Big \rangle_{\widehat{IQ}_{N, \Gamma}^{\rm c} (1,1)}.
\end{align}

The average in (\ref{eq:Bv2}) is an example of the LHS of (\ref{eq:A1}) with 
\begin{equation}
  \label{eq:kac}
  k = -i\Gamma, \qquad a(\vec{r_l}) = -\pi (y_l - \frac{1}{2})^2 + \log 2 \big|\sin \pi \big(x_l + i(y_l - y)  \big)  \big|.
\end{equation}
We seek the corresponding values of $\mu_N$ and $\sigma^2$ on the RHS of (\ref{eq:A1}).

\begin{lemma}
  Let
  \begin{equation}
    \label{eq:M2}
    M_2 = \frac{1}{\pi \Gamma} \big(1- \frac{\Gamma}{4} \big).
  \end{equation}
For the soft cylinder with $\rho_b=N$, $L=W=1$ and $a(\vec{r_l})$ as in (\ref{eq:kac}) with $y<0$ we have 
\begin{align}
  \label{eq:M3}
  \mu_N &= N\pi \big(\frac{5}{12} - y \big) - \pi M_2 + o(1)\\
  \label{eq:M4}
  \sigma_{\rm bulk}^2 &= \frac{1}{2\Gamma} \log \frac{1 -e^{-4\pi(1-y)}}{1 -e^{-4\pi y}} - \frac{4\pi}{3\Gamma} \\
  \label{eq:M5}
  \sigma_{{\rm surface}, 0}^2 &= - \frac{1}{2\Gamma} \log ({1 -e^{-4\pi y}}) \\
  \label{eq:M6}
  \sigma_{{\rm surface}, 1}^2 &= - \frac{1}{2\Gamma} \log ({1 -e^{-4\pi(1-y)}} ).
\end{align}
\end{lemma}

\noindent
Proof. \quad To be able to deduce (\ref{eq:M3}) correct up to the $o(1)$ term, we require the correction term to the global density in the soft cylinder system with $\rho_b = N$, $L=W=1$. This is undertaken in Appendix \ref{AA} where it is shown
\begin{equation}
  \label{eq:M7bis}
  \rho_b^{N , {\rm g}} \big((x,y) \big) = N \chi_{x, y \in [0 ,1]} + \frac{M_2}{4} \big( \delta '' (y) + \delta '' (1-y) \big) + o(1),
\end{equation}
where $M_2$ is given by (\ref{eq:M2}). {{} Note that as for the soft disk case
(\ref{eq:B3}),  the correction term has the simple dependence on $\Gamma$ as given
in (\ref{eq:M2}), and furthermore is supported entirely on the boundary of the
plasma.} Now substituting this and the expression for $a(\vec{r_l})$ (\ref{eq:kac}) in (\ref{eq:A2}), (\ref{eq:M3}) results after an elementary calculation.

The key to deriving (\ref{eq:M4})--(\ref{eq:M6}) from the definitions (\ref{eq:A4}) and (\ref{eq:A7}) is the Fourier expansion
\begin{equation}\label{Yy}
  \log 2 \big|\sin \pi \big(X + i (Y - y) \big)\big| = \pi |Y-y| - \frac{1}{2} \sum_{\substack{p = -\8 \\ p \neq 0}}^\8 \frac{1}{|p|}e^{2\pi i Xp - 2\pi |Y - y||p|}
\end{equation}
(cf.~(\ref{CA3})).
The calculation then becomes elementary. \hfill $\square$

Substituting the result of Lemma 10 in the RHS of (\ref{eq:A1}) with $k$ as in (\ref{eq:kac}) we obtain the large $N$ expansion
\begin{equation}
  \label{eq:M7}
  \big \langle \prod_{l=1}^N e^{-\Gamma \pi (y_l - 1/2)^2} \big|2\sin \big( \pi x_l + \pi i (y - y_l) \big)\big|^\Gamma   \big \rangle_{\widehat{IQ}_{N, \Gamma}^c (1,1)} = \frac{e^{-N\pi \Gamma y}}{(1 - e^{4\pi y})^{\Gamma /2}} e^{5\Gamma N \pi /12 - \pi \Gamma M_2 - 2 \pi \Gamma /3 + o(1)}.
\end{equation}
Furthermore, analogous to (\ref{eq:17.3}) we can make use of  (\ref{eq:d4})  and (\ref{eq:WM2}) to deduce that
\begin{equation}
  \label{eq:M8}
  (N+1) \frac{Q_{N, \Gamma}^c (\sqrt{N}, \sqrt{N})}{Q_{N+1, \Gamma}^{\rm c} (\sqrt{N+1}, \sqrt{N+1})} = \Big(\frac{2\pi}{\sqrt{N+1}}  \Big)^{\Gamma /2} \Big(1 + \frac{1}{N} \Big)^{-N\Gamma /4}e^{-\Gamma (N + \frac{1}{2}) \pi /6} e^{\beta f (\Gamma, 1) + o(1)}.
\end{equation}
Substituting (\ref{eq:M7}) and (\ref{eq:M8}) in (\ref{eq:Bv2}), then replacing $N+1$ by $N$, we obtain the desired large deviation formula.

\begin{prop}
  For the soft cylinder $2d {\rm OCP}$ with $\rho_b=1$, $L = W = \sqrt N$ and corresponding dimensionless free energy per particle $\beta f (\Gamma, \rho_b)$ we have for $y<0$
  \begin{equation}
    \label{eq:21.1}
    \rho_{(1)}^{N , {\rm c}} (\sqrt{N} y) = \Big(\frac{2\pi}{\sqrt{N}}  \Big)^{\Gamma /2} e^{\beta f (\Gamma, 1) + o(1)} \frac{e^{-\Gamma \pi N y^2 + \Gamma \pi y}}{(1- e^{4\pi y})^{\Gamma /2}}.
  \end{equation}
\end{prop}

For $\Gamma = 2$ we can check (\ref{eq:21.1}) against the exact result (\ref{eq:Ay}), upon using the fact that for $\Gamma = 2$, $\beta f (\Gamma, 1) = \frac{1}{2} \log (\frac{1}{2\pi ^2} )$ (recall (\ref{F2a})), and agreement is found.
In Appendix \ref{AA3} theory relating to the term
$o(1)$ in (\ref{eq:17.4}) for $y \to -\infty$ is presented, giving its value as
\begin{equation}\label{4.40a}
\pi(1 - 2 \Gamma)/(6N) + o(1/N)
\end{equation}
in that limit. This furthermore suggests this term for general $y$ to also have leading behaviour
proportional to $1/N$. The validity of (\ref{4.40a}) and the latter claim is verified at $\Gamma = 2$
by inspection of (\ref{eq:Ay}).

The scaled limits of the large deviation formulas, already computed in (\ref{eq:U1}) and (\ref{eq:U2}) in the case $\Gamma = 2$, can now be computed for general $\Gamma > 0$ for both the soft disk and cylinder  {{}(this asymptotic form is also reported in \cite{ZW06}, up to the $O(1)$ term)}.

\begin{cor}
  In an analogous notation to that used on the LHS of (\ref{eq:U1}) and (\ref{eq:U2}) we have 
  \begin{align}
    \label{eq:22.1}
    \lim_{N \to \8} \tilde{\rho}_{(1)}^{N+1, {\rm d}} (\sqrt{N}r)|_{r = 1-y/\sqrt N} & = \lim_{N \to \8} \frac{1}{\pi} \tilde{\rho}_{(1)}^{N, {\rm c}} (\sqrt{N}y) |_{y \mapsto y/\sqrt{\pi N}} \nonumber \\
& = e^{\beta f (\Gamma, 1/\pi)} \frac{e^{-\Gamma y^2}}{(2 |y|)^{\Gamma /2}}.
  \end{align}
\end{cor}

\noindent
Proof. \quad This is immediate from Proposition 2 and 3, together with a simple scaling which shows \cite{AJ81}  (see also (\ref{eq:WM2}) below)
\begin{equation*}
  \beta f (\Gamma, \rho_b) = (1 - \frac{\Gamma}{4} )\log \rho_b + g(\Gamma).
\end{equation*}
\hfill $\square$

In keeping with the discussion of Section \ref{SE} we expect that (\ref{eq:22.1}) is the leading $y \to -\8$ asymptotic form of the edge density profile, for general $\Gamma >0$ and with $\rho_b = 1/\pi$. In addition to the check on this result for $\Gamma = 2$, we see that the leading $y \to -\8$ form of (\ref{eq:22.1}) expanded to first order in $\varepsilon = \Gamma -2$ is precisely that obtained in Lemma 8.


\section*{Acknowledgements}
{{} The work of P.W. and T.C.  was supported by NSF DMS-1156636
and DMS-1206648. The work of P.F. was supported by the Australian Research Council
through the DP `Characteristic polynomials in random matrix theory'.}
G.T. acknowledges financial support from Fa\-cultad de Ciencias, Uniandes.

\appendix
\makeatletter
\def\@seccntformat#1{Appendix\ \csname the#1\endcsname\quad}
\makeatother

\section{}\label{AA1}
\setcounter{equation}{0}
The purpose of this appendix is to derive (\ref{V1}).

According to the definitions, for general $\Gamma$ in soft cylinder geometry
\begin{align}\label{ST1}
\langle U_1 \rangle^{\rm c} & = - {1 \over 2} \int_0^W dx_1 \int_0^W dx_2 \int_{-\infty}^\infty dy_1   \int_{-\infty}^\infty dy_2 \,
\log 2 \Big | \sin {\pi ((x_1 - x_2) + i (y_1 - y_2)) \over W} \Big | \nonumber \\
& \quad \times \rho_{(2)}(\vec{r}_1,\vec{r}_2),
\end{align}
where $\vec{r}_j = (x_j,y_j)$. Generalizing (\ref{F12}), we know that for $\Gamma = 2$ 
\cite{CFS83}
$$
\rho_{(2)}(\vec{r}_1,\vec{r}_2) = \rho_{(1)}(\vec{r}_1)   \rho_{(1)}(\vec{r}_2) +   \rho_{(2)}^T(\vec{r}_1,\vec{r}_2)
$$
where $ \rho_{(1)}(\vec{r})$ is given by (\ref{F12}) and
\begin{align*}
\rho_{(2)}^T(\vec{r}_1,\vec{r}_2) & = - {2 \rho_b \over W^2} e^{- \pi (y_1 - y_2)^2}
\sum_{q_1 = 0}^{N-1} \exp \Big \{ - 2 \pi \rho_b \Big ( {y_1 + y_2 \over 2} - {q_1 + 1/2 \over W \rho_b} \Big )^2 + 2 \pi i q_1 {(x_1 - x_2) \over W}  \Big \} \nonumber \\
& \quad \times
\sum_{q_2 = 0}^{N-1} \exp \Big \{ - 2 \pi \rho_b \Big ( {y_1 + y_2 \over 2} - {q_2 + 1/2 \over W \rho_b} \Big )^2 - 2 \pi i q_2 {(x_1 - x_2) \over W}  \Big \}
\end{align*}
(the case $l=2$ of (\ref{CA1})).
Our task then is to compute some explicit multiple integrals.

Making use of the Fourier expansion (\ref{Yy}), elementary calculations show
\begin{align}\label{Fu1}
&  - {1 \over 2} \int_0^W dx_1 \int_0^W dx_2 \int_{-\infty}^\infty dy_1   \int_{-\infty}^\infty dy_2 \,
\log 2 \Big | \sin {\pi ((x_1 - x_2) + i (y_1 - y_2)) \over W} \Big |   \rho_{(2)}^T(\vec{r}_1,\vec{r}_2)  \nonumber \\
& \quad = - {1 \over \pi} \sum_{l=1}^{N-1} {N - l \over l}
\int_{-\infty}^\infty dy_1 \, e^{-y_1^2} \int_{y_1 + \sqrt{2 \pi/\rho_b} l / W}^\infty dy_2 \, e^{-y_2^2} + {N \over W \sqrt{\rho_b}}
\end{align}
and
\begin{align}\label{Fu2}
&  - {1 \over 2} \int_0^W dx_1 \int_0^W dx_2 \int_{-\infty}^\infty dy_1   \int_{-\infty}^\infty dy_2 \,
\log 2 \Big | \sin {\pi ((x_1 - x_2) + i (y_1 - y_2)) \over W} \Big |   \rho_{(1)}(\vec{r}_1)  \rho_{(1)}(\vec{r}_2)  \nonumber \\
& \quad = - {\pi \over W^2 \rho_b} \sum_{l=1}^{N-1} (N-l) l + {2 \over W}
\sum_{l=1}^{N-1} (N-l) {l \over W \rho_b} 
\int_{-\infty}^\infty dy_1 \, e^{-y_1^2} \int_{y_1 + \sqrt{2 \pi/\rho_b} l / W}^\infty dy_2 \, e^{-y_2^2}   \nonumber  \\
& \qquad - {1 \over W \sqrt{\rho_b}} \sum_{l=1}^{N-1} (N-l) e^{-\pi l^2/\rho_b W^2}.
\end{align}
This reduces our task to analyzing certain one-dimensional sums in the large $N$ limit.

The first sum in (\ref{Fu2}) is elementary, and we have 
\begin{equation}\label{Sw}
\sum_{l=1}^{N-1} (N-l) l = N(N^2 - 1)/6. 
\end{equation}
For the remaining sums, the leading and first order
correction for large $N$ can be obtained by making use of the trapezoidal rule
\begin{equation}\label{Fu3}
\sum_{k=1}^N f(k h) = {1 \over h} \int_0^{Nh} f(x) \, dx - \Big ( {f(0) - f(Nh) \over 2} \Big ) + \mathcal O( h^2).
\end{equation}
In this regards, the portion of the first summation in (\ref{Fu1}),
$$
- {N \over \pi} \sum_{l=1}^{N-1} {1\over l}
\int_{-\infty}^\infty dy_1 \, e^{-y_1^2} \int_{y_1 + \sqrt{2 \pi/\rho_b} l / W}^\infty dy_2 \, e^{-y_2^2} 
$$
requires preliminary manipulation, since a literal application of (\ref{Fu3}) is not possible. This is due to the corresponding $f(x)$ not being
integrable about $x=0$. Thus we write
\begin{align}\label{Fu6}
& \sum_{l=1}^{N-1} {1 \over l} 
\int_{-\infty}^\infty dy_1 \, e^{-y_1^2} \int_{y_1 + \sqrt{2 \pi/\rho_b} l / W}^\infty dy_2 \, e^{-y_2^2}  \nonumber \\
& \quad = \sum_{l=1}^K {1 \over l} \Big \{
\int_{-\infty}^\infty dy_1 \, e^{-y_1^2}  \int_{y_1 + \sqrt{2 \pi/\rho_b} l / W}^\infty dy_2 \, e^{-y_2^2}  - \int_{-\infty}^\infty dy_1 \, e^{-y_1^2}
 \int_{y_1}^\infty   \, e^{-y_2^2}  \Big \} \nonumber \\
& \qquad + 
\Big (  \int_{-\infty}^\infty dy_1 \, e^{-y_1^2}
 \int_{y_1}^\infty   \, e^{-y_2^2}  \Big ) \sum_{l=1}^K {1 \over l}   \nonumber \\
& \qquad +  \sum_{l=K+1}^N {1 \over l} 
\int_{-\infty}^\infty dy_1 \, e^{-y_1^2} \int_{y_1 + \sqrt{2 \pi/\rho_b} l / W}^\infty dy_2 \, e^{-y_2^2} ,
\end{align}
where $K = \Big [ W \sqrt{2 \pi \over \rho_b} \Big ]$. 

With $H_K$ denoting the harmonic numbers, it is a standard result that
\begin{equation}\label{Fu7}
\sum_{l=1}^K {1 \over l} =: H_K = \log K + {\bf C} + {1 \over 2K} + \mathcal O \Big ( {1 \over K^2} \Big ).
\end{equation}
The remaining sums in (\ref{Fu6}) can all be analyzed using (\ref{Fu3}).
Doing this and combining with (\ref{Fu7}) shows
\begin{align}\label{Eu1}
& - {1 \over \pi} \sum_{l=1}^{N-1} {1 \over l} 
\int_{-\infty}^\infty dy_1 \, e^{-y_1^2} \int_{y_1 + \sqrt{2 \pi/\rho_b} l / W}^\infty dy_2 \, e^{-y_2^2}  \nonumber \\
& \quad = - {N \over 2} \log \Big ( \sqrt{\rho_b \over 2} {W \over 2} \Big ) - {N {\bf C} \over 4} - {N \over 2 W \sqrt{\rho_b}} +
{\sqrt{\rho_b} \over 2 \pi} W + \mathcal O(1)
\end{align}
and
\begin{align}\label{Eu2}
&  {2 \over W}
\sum_{l=1}^{N-1} (N-l) {l \over W \rho_b} 
\int_{-\infty}^\infty dy_1 \, e^{-y_1^2} \int_{y_1 + \sqrt{2 \pi/\rho_b} l / W}^\infty dy_2 \, e^{-y_2^2}   
 - {1 \over W \sqrt{\rho_b}} \sum_{l=1}^{N-1} (N-l) e^{-\pi l^2/\rho_b W^2} \nonumber \\
& \quad = \Big ( {N \over 4} - {1 \over 3 \pi} W \sqrt{\rho_b} \Big ) +
\Big ( - {N \over 2} - {N \over 2 W \sqrt{\rho_b}} + {\sqrt{\rho_b} W \over 2 \pi} \Big ) +  \mathcal O(1).
\end{align}
Substituting (\ref{Eu1}) in (\ref{Fu1}), (\ref{Eu2}) and (\ref{Sw}) in (\ref{Fu2}), and using these results to evaluate the RHS of (\ref{ST1}) gives (\ref{V1}).
\section{}\label{AA}
\setcounter{equation}{0}
Consider the soft cylinder with leading order density profile in the $y$-direction $\tilde{\rho_b} = \rho_b \chi _{0 < y < W} $. For large $W$, $n \in \mathbb{Z}^+$, we see that
\begin{equation*}
  \int_{-\8}^\8 (y - W/2)^{2n} (\rho_{(1)}^{N,{\rm c}}(y) - \tilde{\rho_b}) dy \sim \frac{2n (2n-1)}{2} \Big( \frac{W}{2} \Big)^{2n-2} \tilde{M_2}, \quad  \tilde{M_2} := \int_{-\8}^\8 y^2 (\rho_{(1)}^{N,{\rm c}}(y) - \tilde{\rho_b}) dy.
\end{equation*}
A readily verifiable consequence is that to leading order
\begin{equation}
  \label{eq:rr}
  \rho_{(1)}(y) - \tilde{\rho_b} = \frac{\tilde{M_2}}{4} \big(\delta '' (W-y) + \delta '' (y)  \big).
\end{equation}
We observe that the RHS of (\ref{eq:rr}), multiplied by the measure $dy$, is independent of $W$ if we scale $y \mapsto Wy$, $x \mapsto Wx$, $\frac{1}{W} \tilde{M_2} \mapsto M_2$, where
\begin{equation}
  \label{eq:rr1}
  {M_2} := \int_{-\8}^\8 y^2 (\rho_{(1)}^{N,{\rm c}}(y) |_{L=W=1}- N \chi_{0<y<1}) dy
\end{equation}
Thus (\ref{eq:M7bis}) follows, provided we can show that $M_2$ has the evaluation (\ref{eq:M2}).

For this latter task we observe from the explicit formula for the partition function implied by (\ref{eq:d4}) that
\begin{equation*}
  W \frac{\partial}{\partial W} \log Z_{N, \Gamma} (W,L) = \frac{-\pi \Gamma \rho_b W^2}{3}N + \Gamma \pi \rho_b \bigg \langle \sum_{l=1}^N y_l^2  \bigg \rangle_{\widehat{IQ}_{N, \Gamma}(W,L)}.
\end{equation*}
Changing variables $x_l \mapsto x_l / L$, $y_l \mapsto y_l / L$ and setting $W=L$ this reads 
\begin{align}
  \label{eq:WM}
  W \frac{\partial}{\partial W} \log Z_{N, \Gamma} (W,L)|_{W=L} & = \frac{-\pi \Gamma N^2}{3} + \Gamma \pi N \bigg \langle \sum_{l=1}^N y_l^2  \bigg \rangle_{\widehat{IQ}_{N, \Gamma}(1,1)} \nonumber \\
& = \int_{-\8}^\8 y^2 (\rho_{(1)}^{N,{\rm c}} (y) |_{L=W=1} - N \chi_{0<y<1})dy =: M_2.
\end{align}
Thus we seek an independent computation of the LHS of (\ref{eq:WM}).

To provide such a computation, we first observe 
\begin{equation}
  \label{eq:WM1}
  W \frac{\partial}{\partial W} = -\rho_b \frac{\partial}{\partial \rho_b}.
\end{equation}
Next we note that scaling in disk geometry together with the expected universality of the leading large $N$ behaviour of the partitions in disk and cylinder geometries implies that for large $N$
\begin{equation}
  \label{eq:WM2}
  Z_{N, \Gamma} (W,L) |_{\rho_b = N/{WL}} \sim e^{N(\Gamma / 4 - 1)\log \rho_b + Ng(\Gamma) + \mathcal{O}(\sqrt N)}
\end{equation}
for some $g(\Gamma)$. Substituting (\ref{eq:WM1}) and (\ref{eq:WM2}) in the LHS of (\ref{eq:WM}) gives (\ref{eq:M2})

\section{}\label{AA3}
\setcounter{equation}{0} In this appendix, we study the behavior of
the density in the cylinder when $y\to-\infty$ for finite $N$ and $W$,
when $\Gamma/2$ is an integer. We will consider first $N$ and $W$ as
independent variables. Let $\tilde{W}=\rho_b W^2$ and
$\tilde{y}=\rho_b W y$ be the rescaled lengths by the characteristic
length $1/(\rho_b W)$. Considerations leading to the configuration
integral (\ref{k10c}) can be extended to obtain the density profile~\cite{SWK04}
\begin{equation}
  \label{eq:rho_cyl_G468}
  \rho_{(1)}^{N,c}(y)=
  \rho_b \sqrt{\frac{\Gamma}{\tilde{W}}}
  \
  \sum_{l=0}^{(N-1)\Gamma/2}
  a_l^{\rm c} \exp\left[-\frac{2\pi\Gamma}{\tilde{W}}
    \left(
      \tilde{y}-N+\frac{1}{2}+\frac{2l}{\Gamma}
    \right)^2
  \right]
\end{equation}
with
\begin{equation}
  \label{eq:al}
  a_l^{\rm c}=\frac{1}{Q_{N,\Gamma}^{\rm c *}}
  \sum_{\mu\, |\, l\in\mu}
    {(c_\mu^{(N)}(\Gamma/2))^2 \over \prod_i m_i!} e^{\pi \Gamma 
      \sum_{j=1}^N (2\mu_j/\Gamma + 1/2)^2/\tilde{W}},
\end{equation}
where the sum runs over all partitions which include $l$. If
$\tilde{y}\to-\infty$, then
\begin{equation}
  \label{eq:rho_cyl_prelimG468}
  \rho_{(1)}^{N,c}(y) \underset{y\to-\infty}{\sim}
  \rho_b \sqrt{\frac{\Gamma}{\tilde{W}}}
  \,
  e^{-\pi\Gamma (\tilde{y}-1/2)^2/\tilde{W}}
  a_{(N-1)\Gamma/2}^{\rm c}
  \ .
\end{equation}
To compute $a_{(N-1)\Gamma/2}^{\rm c}$, one needs to consider in (\ref{eq:al})
all the partitions $\mu$ with $c_{\mu}^{(N)}(\Gamma/2)\neq0$ and
$\mu_1=(N-1)\Gamma/2$. The partition
$\tilde{\mu}=(\mu_2,\mu_3,\ldots,\mu_{N})$ is a partition of
$\Gamma(N-1)(N-2)/4$ with $\Gamma (N-2)/2 \leq \mu_2 \leq
\cdots \leq {\mu}_N$, and due to a factorization property
satisfied by the coefficients of the partitions~\cite{BR09}, one has
\begin{equation}
  \label{eq:cmufactor}
  c_{((N-1)\Gamma/2,\tilde{\mu})}^{(N)}(\Gamma/2) =    
  c_{\tilde{\mu}}^{(N-1)}(\Gamma/2)
  \,.
\end{equation}
Therefore $\tilde{\mu}$ corresponds to a partition for a system with
$N-1$ particles (this is not surprising as taking $y \to \infty$ effectively removes
that particle; see a similar argument in \cite{FM09}). Then
\begin{equation}
  \label{eq:aQQ}
  a_{(N-1)\Gamma/2}^{\rm c}=
  \frac{Q_{N-1,\Gamma}^{\rm c *}(\tilde{W})}{Q_{N,\Gamma}^{\rm c *}(\tilde{W})}
  \, e^{\pi\Gamma(N-1/2)^2}
  \,,
\end{equation}
and using~(\ref{eq:Fcyl}), this leads to
\begin{eqnarray}
  \rho_{(1)}^{N,c}(y) &\underset{y\to-\infty}{\sim}&
  \rho_b \left(\frac{2\pi}{\sqrt{\tilde{W}}}\right)^{\Gamma/2}
  e^{-\pi\Gamma (\tilde{y}^2-\tilde{y})/\tilde{W}}
  \nonumber\\
  &&
  \times 
  \exp\left[\beta[F^{c}_{N,\Gamma}(\tilde{W})-F^{c}_{N-1,\Gamma}(\tilde{W})]
    -\left(1-\frac{\Gamma}{4}\right)\log\rho_b
    -\frac{\pi\Gamma}{3\tilde{W}}\right].
  \label{eq:rhocyl-asymptG468}
\end{eqnarray}
Now, consider the limit $N\to\infty$, and
$\tilde{W}\to\infty$, but with $N$ and $\tilde{W}$ independent. Using
the universal properties of the free energy (\ref{eq:Fexpans}), we
have
\begin{eqnarray}
  \rho_{(1)}^{N,c}(y) &\underset{y\to-\infty}{\sim}&
  \rho_b \left(\frac{2\pi}{\sqrt{\tilde{W}}}\right)^{\Gamma/2}
  e^{-\pi\Gamma (\tilde{y}^2-\tilde{y})/\tilde{W}} 
  \nonumber\\
&&\times  \exp\left[\beta f(\Gamma,1)
   + (1-2\Gamma) \frac{\pi}{6\tilde{W}}+o(1/N)+o(1/\tilde{W})
   \right]
  \,.
  \label{eq:rhocyl_scalededgeG468}
\end{eqnarray}
Notice that in the difference
$F^{c}_{N,\Gamma}(\tilde{W})-F^{c}_{N-1,\Gamma}(\tilde{W})$, as
$\tilde{W}$ is kept fixed, the surface tension terms in
(\ref{eq:Fexpans}) cancel out, leading to a next order correction of
order ${\cal O}(1/N)$ instead of a naively expected ${\cal
  O}(1/\sqrt{N})$.  In the scaled edge $\tilde{W}=N\to\infty$ and
$\tilde{y}\mapsto N y$ this can be compared to (\ref{eq:21.1}). Indeed
if one takes $y\to-\infty$ in (\ref{eq:21.1}), then
(\ref{eq:rhocyl_scalededgeG468}) is recovered. The $o(1)$ term 
in~(\ref{eq:21.1}) for $y \to - \infty$ should be (\ref{4.40a}).

\begin{figure}[tbp]
  \centering
  \includegraphics[width=10cm]{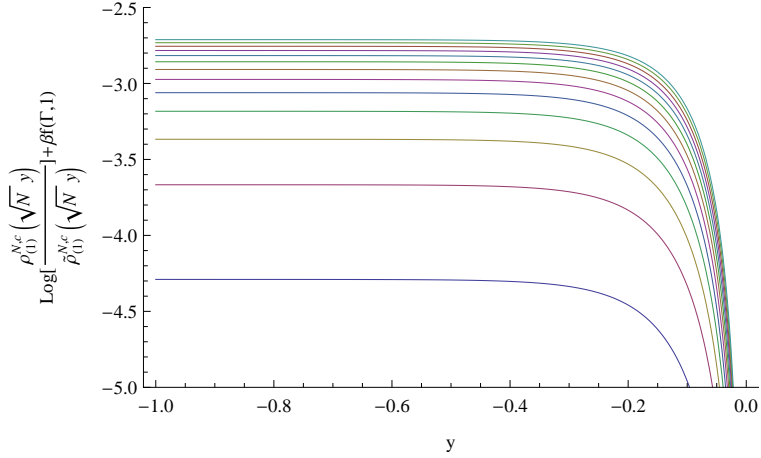}
  \caption{Exact numerically computed density profile in the soft
    cylinder compared to the scaled form (\ref{eq:21.1}). From bottom
    to top, $W^2=N=2,3,4,5,6,7,8,9,10,11,12,13,14$.}
  \label{fig:rhoscaledG4}
\end{figure}

\begin{figure}[tbp]
  \centering
  \includegraphics[width=10cm]{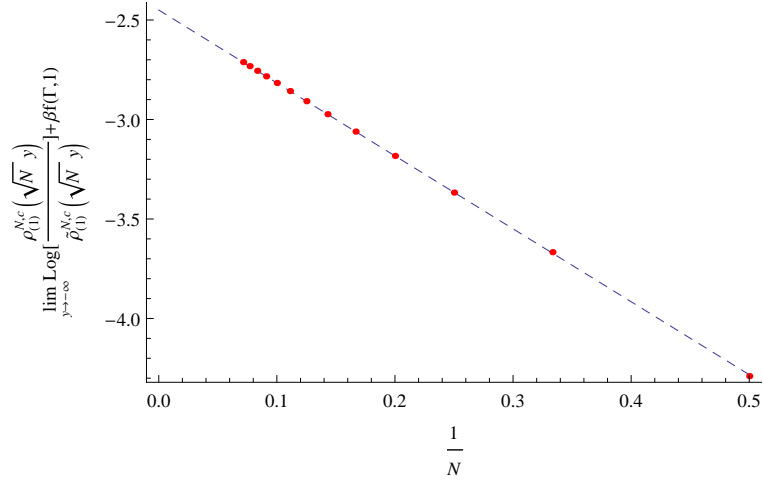}
  \caption{Numerical value of the LHS of
    (\ref{eq:limitrhorhoasym_cyl}) as a function of $N$ (red dots) and a
    linear regression done with values of $N>7$ (blue dashed line).}
  \label{fig:linreg}
\end{figure}

As an illustration of the results, for $\Gamma=4$,
Figure~\ref{fig:rhoscaledG4} shows a plot of the numerically computed
$$\log(\rho_{(1)}^{N,c}(\sqrt{N}y)/\tilde{\rho}_{(1)}^{N,c}(\sqrt{N}y))
+\beta f(\Gamma,1),$$
 for various values of $N=\tilde{W}$ confirming
the expected behavior as $y\to-\infty$. 
{{} In the plot, $\tilde{\rho}_{(1)}^{N,c}$ denotes the right
  hand side of (\ref{eq:21.1}).
}
In Figure~\ref{fig:linreg}, the
value of the limit of
$\log(\rho_{(1)}^{N,c}(y)/\tilde{\rho}_{(1)}^{N,c}(\sqrt{N}y))+\beta
f(\Gamma,1)$ as $y\to-\infty$ is plotted against $1/N$, showing indeed
a linear behavior as expected
\begin{equation}
  \label{eq:limitrhorhoasym_cyl}
 \lim_{y\to-\infty} \log\frac{\rho_{(1)}^{N,c}(y)}{\tilde{\rho}_{(1)}^{N,c}(\sqrt{N}y)e^{-\beta f(\Gamma,1)}}
  = \beta f(\Gamma,1) +\frac{\pi}{6}(1-2\Gamma) \frac{1}{N}
  +o(1/N)
  \,.
\end{equation}

Very similar figures are obtained for $\Gamma=6$ and 8 (not shown).
Doing a numerical regression of Figure~\ref{fig:linreg} provides an
alternative way to obtain numerically $g(\Gamma)=\beta f(\Gamma,1)$,
and verify the $1/N$ finite size correction. Table
\ref{tab:reg-rho-rhotilde} shows the values obtained for $g(\Gamma)$
and the $1/N$ correction for $\Gamma=4,$ 6, 8, and compares them to the
estimations of free energy per particle on the sphere~\cite{TF99} and
the expected value $\pi(1-2\Gamma)/6$ of the $1/N$ correction. As this
method for estimating the free energy per particle relies on fitting
an expression with $1/N$ corrections, it seems as equally reliable as
the one used in \cite{TF99} for the 2dOCP on the sphere when the
universal $\log N$ correction is subtracted to the free energy.

\begin{table}[tbp]
  \centering
  \begin{tabular}{|c||c|c|c|}
    \hline
    $\Gamma$ & 4 & 6 & 8 \\
    \hline 
    $g$ (cylinder) & -2.449893 & -3.5168 & -4.641 \\
    $g$ (sphere)   & -2.449884 & -3.5175 & -4.639 \\
    Relative 
    difference   & 0.00037\%   & 0.020\% & 0.04\% \\
    \hline
    $1/N$ correction   & -3.665103757  & -5.767068913 &  -7.842621261 \\
    Exact value: $\pi(1-2\Gamma)/6$ & -3.665191429  & -5.759586532 &  -7.853981634 \\ 
    Relative difference 
     & 0.00239\% & 0.130\% & 0.145\%     
    \\
    \hline
  \end{tabular}
  \caption{Estimation of the free energy $g(\Gamma)=\beta f(\Gamma,1)$ per particle obtained
    from~(\ref{eq:limitrhorhoasym_cyl}). Linear regressions where done
    with $8\leq N\leq14$ for $\Gamma=4$ and 6, and with $7\leq N \leq
    11$ for $\Gamma=8$.}
  \label{tab:reg-rho-rhotilde}
\end{table}

Similar considerations can be done for the soft disk. The density
profile is~\cite{TF99}
\begin{equation}
  \label{eq:rhosoftdisk}
  \rho_{(1)}^{\rm d}(r)=(\Gamma/2) \rho_b  e^{-\pi\Gamma \rho_b r^2/2} 
  \sum_{l=0}^{(N-1)\Gamma/2} a_l^{\rm d} \, (\Gamma \pi\rho_b r^2/2)^{l}
  \,,
\end{equation}
with
\begin{equation}
  \label{eq:alsoftdisk}
  a_l^{\rm d}=\frac{N!\pi^{N}}{Q_{N,\Gamma}^{\rm d}(\rho_b)\, l!}
  \sum_{\mu\,|\,l\in\mu}
  \frac{(c_\mu^{(N)}(\Gamma/2))^2}{\prod_{i}m_i!}
  \prod_{j=1}^{N-1} \mu_j!
\,.
\end{equation}
The leading behavior of the density as $r\to\infty$ is given by
\begin{equation}
  \label{eq:rhoasymdisk}
  \rho_{(1)}^{\rm d}(r)\underset{y\to-\infty}{\sim}
  (\Gamma/2)\,\rho_b\,  e^{-\pi\Gamma \rho_b r^2/2}\,
  a_{(N-1)\Gamma/2}^{\rm d}
  (\pi\rho_b\Gamma r^2/2)^{(N-1)\Gamma/2}
  \,.
\end{equation}
Again, the coefficient $a_{(N-1)\Gamma/2}^{\rm d}$ is related to the
ratio of two partition functions with $N$ and $N-1$ particles
\begin{equation}
  \label{eq:adiskQnn-1}
  a_{(N-1)\Gamma/2}^{\rm d}= N \pi 
  \frac{Q_{N-1,\Gamma}^{\rm d}(\rho_b)}{Q_{N,\Gamma}^{\rm d}(\rho_b)} 
  \,.
\end{equation}
Using (\ref{k1}), we find
\begin{multline}
  \label{eq:rhodiskrinf}
  \rho_{(1)}^{\rm d}(r)\underset{y\to-\infty}{\sim}
  \frac{\pi\rho_b}{N^{\Gamma/4}} 
  \exp\left[-\frac{N\Gamma}{2}\left(\frac{\pi\rho_b r^2}{N}-1\right)
  \right]
    \left(
      \frac{\pi\rho_b r^2}{N}
    \right)^{(N-1)\Gamma/2}
    \\
  \times  \exp
    \left[
      \beta f(\Gamma,\rho_b)-\left(1-\frac{\Gamma}{4}\right)
      \log(\pi\rho_b)
      +\frac{\beta \mu(\Gamma,\rho_b)\sqrt{\pi}}{\sqrt{\rho_b N}}
      +(1-\Gamma)\frac{1}{12N}+o(1/N)
    \right]
\end{multline}
In the scaled edge, with $r\mapsto \sqrt{N}r$ and $\rho_b =1/\pi$,
taking $r\to\infty$ in (\ref{eq:17.4}) reproduces
(\ref{eq:rhodiskrinf}), but here the $o(1)$ has non zero ${\cal
  O}(1/\sqrt{N})$ corrections --- except for $\Gamma = 2$ when $\beta \mu(\Gamma,\rho_b)$ vanishes --- as opposed to the soft cylinder geometry.

\section{} \label{AA5}

In this appendix we present a detailed derivation of the exterior asymptotes of $A(y)$ as in (\ref{eq:7.0}). From Proposition 1, \(A(y)\) can be written as a sum of four terms, each of which is analyzed separately below.

\begin{lemma}
The asymptotic expansion of $A_{1}(y)$ outside the droplet is

 \begin{equation*}
    A_{1}(y) \mathop{\sim}_{y \to -\8}  - \frac{1}{8} \frac{e^{- 2y^{2}}}{\sqrt{2\pi} |y|} + \frac{1}{16\pi \sqrt{2}} \frac{e^{- 2y^{2}}}{ y^{2}} +\frac{1}{32 \sqrt{2\pi} |y|^{3}} e^{- 2y^{2}} +  \mathcal{O}(y^{-4}e^{- 2y^{2}}).
  \end{equation*}

\end{lemma}
Proof. This asymptotic expansion can be obtained by differentiating $A_{1}(- |y|)$ with respect to $|y|$, and integrating from $|y|$ to $\infty$, with the result

\begin{align*}
A_{1}(- |y|) 
& = - \frac{1}{8} {\rm erfc}(\sqrt{2}|y|) + \frac{1}{2 \sqrt{6\pi}} \int_{|y|}^{\infty} dt \,
e^{- 2t^{2}/3} {\rm erfc}(2 t/\sqrt{3}), \\
&= - \frac{1}{8} {\rm erfc}(\sqrt{2}|y|) + \frac{1}{16\pi \sqrt{2}} \frac{e^{- 2y^{2}}}{ y^{2}} + \mathcal{O}(y^{-4}e^{- 2y^{2}}).
\end{align*}

The second line is obtained by expanding the complementary error function in the integrand
for large $t$, and integrating by parts. The result follows by keeping the next to leading order term in the large \(|y|\) expansion of the first term.
\hfill $\square$
\\

This can be used to show the following.
\begin{lemma}
The leading order asymptote of $A_{2}(y)$ outside the droplet is 

\begin{equation*}
A_{2}(y)  \mathop{\sim}_{y \to -\8} \mathcal{O}\left( y^{-2} e^{- 2y^{2}}\right).
\end{equation*}
\end{lemma}

\noindent
Proof. \quad Applying a sequence of integration by parts, we can rewrite $A_{2}(y)$ in terms of $A_{1}(y)$ as
\begin{equation*}
A_{2}(y)  = -\frac{1}{4 \sqrt{2\pi}} y e^{- 2y^{2}} + \frac{1}{4 \sqrt{2}\pi} e^{- 2y^{2}}  + \frac{3}{4 \sqrt{6\pi}} y e^{- 2y^{2}/3} {\rm erfc}(-2 y/\sqrt{3})+ \left( \frac{1}{2} +2 y^{2}\right) A_{1}(y).
\end{equation*}

Using the asymptotic expansion for $A_{1}(-|y|)$ above, only the term with a pre-exponential factor of \(\mathcal{O}(y^{-2})\) remains. 
\hfill $\square$

\smallskip

Next, we consider the leading asymptote of $A_{3}(y)$. This follows by a straightforward expansion for large \(-y \gg 1\).

\begin{lemma}
The leading asymptote of \(A_{3}(y)\) outside the droplet is
\begin{equation*}
A_{3}(y)  \mathop{\sim}_{y \to -\8}  \frac{1}{2 \sqrt{2\pi}} |y| e^{- 2y^{2}} + \mathcal{O} \left( y^{- 3} e^{- 2y^{2}}\right).
\end{equation*}
\end{lemma}

\noindent
Proof. \quad After replacing the error functions appearing in \(A_{3}(y)\) with their large \(|y|\) asymptotic expansions, this result follows by straightforward algebra.

\begin{lemma}
The asymptote of $A_{4}(y)$ outside the droplet is
  \begin{equation*}
A_{4}(y)  \mathop{\sim}_{y \to -\8}   \frac{1}{4 \sqrt{2\pi}}  \frac{\log |y|}{|y|} e^{- 2y^{2}}   + \frac{1}{4\sqrt{2\pi}} \left( \frac{{\bf C}}{2}+ \log 2\right) \frac{e^{- 2y^{2}}}{ |y|} + \mathcal{O} \left(y^{- 2} e^{- 2y^{2}}\right).
  \end{equation*}
\end{lemma}

\noindent
Proof. \quad Using the fact that the integrand is symmetric in its arguments $t_{1}$ and $t_{2}$, we can rewrite \(A_{4}(y)\) for \(y < 0\) as
\begin{align*}
A_4(-|y|) = & -\frac{1}{\sqrt{2\pi}}\int_0^\infty dt_1 \int_0^{\infty}dt_2 \frac{e^{-2(t_1+|y|)^2}}{t_1-t_2} \bigg( {\rm erf} (t_1-t_2) + {\rm erf} \big(\sqrt{2}(t_2+|y|) \big)\bigg).
\end{align*}
After a change of variables, 
\begin{align*}
A_4(-|y|) & =  -\frac{1}{\sqrt{2\pi}}\int_{|y|}^\infty dt_1 \int_{|y|}^{\infty}dt_2 \frac{e^{-2 t_{1}^2}}{t_1-t_2} \bigg( {\rm erf} (t_1-t_2) + {\rm erf} \big(\sqrt{2}t_2 \big)\bigg)\\
& \sim -\frac{1}{\sqrt{2\pi}}\int_{|y|}^\infty dt_1 \int_{|y|}^{\infty}dt_2 \frac{e^{-2 t_{1}^2}}{t_1-t_2} \bigg( {\rm erf} (t_1-t_2) +1 - \frac{e^{- 2t _{2}^{2}}}{\sqrt{2\pi} t_{2}}\bigg).
\end{align*}
Integrating over \(t_{2}\) this reads
\begin{align*}
A_{4}(-|y|) \sim - \frac{1}{\sqrt{2\pi}} \int_{|y|}^{\infty} dt_{1} e^{- 2 t_{1}^{2}} \bigg(\log ( t_{1} - |y|) + \frac{{\bf C}}{2} + \log 2 - \int_{0}^{|y| - t_{1}} \frac{{\rm erf}(t)}{t} dt\bigg).
\end{align*}

We can expand the last integral as an asymptotic  series in $(|y| -
t_{1})$. The leading term is $2 (|y| - t_{1})/\sqrt{\pi}$, which, upon
integrating with respect to $t_{1}$, becomes
\begin{equation}
\label{d1}
\int_{|y|}^{\infty} dt_{1} e^{- 2 t_{1}^{2}} (|y| - t_{1}) =\mathcal{O}\left( y^{- 2} e^{- 2 y^{2}} \right).
\end{equation}
The first term in parentheses can be similarly developed as an asymptotic series. A change of variables \(x = t_{1} - |y|\), followed by a rescaling \(x = \xi/|y|\), makes the Gaussian factor \(\exp\left(- 2(x + |y|)^{2}\right) = e^{- 2 y^{2}} \exp\left( -  \frac{2\xi^{2}}{y^{2}} - 4 \xi\right)\). After a Laurent expansion in \((\xi/y)^{2}\), the integral becomes
\begin{align}
\label{d2}
\int_{|y|}^{\infty} dt_{1}e^{- 2t_{1}^{2}} \log(t_{1}-|y|) & =  \frac{e^{- 2y^{2}}}{y} \int_{0}^{\infty}d\xi \log(\xi/y) e^{- 4 \xi}
\left(1 - \frac{2 \xi^{2}}{y^{2}} + \frac{2 \xi^{4}}{y^{4}} + ... \right)\nonumber\\
& = - \frac{\log |y|}{4|y|}e^{- 2y^{2}} -2 \left(  \frac{{\bf C}}{2} +  \log(2)\right)
\frac{1}{4 |y|}e^{- 2y^{2}} + \mathcal{O}\left( \frac{\log y}{y^{3}} e^{- 2y^{2}}\right).
\end{align}

The next term can be evaluated easily and its large distance asymptote reads
\begin{equation}
\label{d3}
\int_{|y|}^{\infty} dt_{1} e^{- 2t_{1}^{2}} \left( \frac{{\bf C}}{2} + \log 2\right) \sim \left( \frac{{\bf C}}{2} + \log 2\right)  \frac{e^{- 2y^{2}}}{4 |y|}.
\end{equation}

Combining (\ref{d1}), (\ref{d2}), and (\ref{d3}) gives the stated asymptote. \hfill $\square$
\\

This exhaustive analysis demonstrates that the leading asymptote outside indeed arises from \(A_{3}(y)\), and moreover
\begin{equation}
A(y)  \mathop{\sim}_{y \to -\8}     \frac{1}{2 \sqrt{2\pi}} |y| e^{- 2y^{2}} + 
 \frac{1}{4 \sqrt{2\pi}}  \frac{\log |y|}{|y|} e^{- 2y^{2}}   + \frac{1}{4\sqrt{2\pi}} \left( \frac{{\bf C}-1}{2}+ \log 2\right) \frac{e^{- 2y^{2}}}{ |y|} +\mathcal{O} \left(y^{- 2} e^{- 2y^{2}}\right).
 \end{equation}

 \section{}\label{AA6}
 
In this appendix we present a more detailed proof of equation (\ref{eq:ode1}) in Lemma \ref{lem:ode}. A direct computation of the LHS for the antisymmetric parts of \(A_{1}(y)\), \(A_{2}(y)\) and \(A_{3}(y)\) gives 
\begin{align}\label{E5}
 \sum_{i = 1}^{3} \left(A_{i,a}''(y) + 4 y A_{i,a}'(y)\right) = \left(2y^{2}+\frac{1}{2}\right)\left( {\rm erf}(\sqrt{2/3}y)-{\rm erf}(\sqrt{2}y)\right)  + \frac{ \sqrt{6}}{ \sqrt{\pi}} y e^{-2 y^{2}/3}.
\end{align}

For \(A_{4}(y)\), we write the LHS as \(e^{-2y^{2}}\partial_{y}\left( e^{2 y^{2}} \partial_{y} A_{a,4}(y)\right)\), and carry out the operations in the sequence implied. First, the antisymmetric part must be written in a suitable form. Taking advantage of the symmetry of the integrand and changing variables, the double integral can be written as

\begin{align*}
  A_4(y) & = - \frac{1}{2 \sqrt{2\pi}} \int_{-y}^{\infty} \int_{-y}^{\infty} dt_{1} dt_{2} F(t_{1},t_{2}),\\
F(t_{1},t_{2}) &= \frac{1}{t_1-t_2}  \bigg(e^{-2t_1^2} \Big({\rm erf} (t_1-t_2) + {\rm erf} \big(\sqrt{2}t_2 \big) \Big)\\
&\qquad\qquad  \qquad + e^{-2 t_2^2} \Big({\rm erf} (t_1-t_2) - {\rm erf} \big(\sqrt{2} t_1 \big)  \Big) \bigg).
\end{align*}

Using the fact that \(\lim_{y \to \8}A_{4}(y) = 0\), this can be written equivalently as
\begin{align*}
 A_{4}(y) & =   \frac{1}{2\sqrt{2\pi}} \int_{-\infty}^{\infty}dt_{1}
\int_{-\infty}^{-y} \, dt_{2} \,F(t_{1},t_{2})  + \frac{1}{2\sqrt{2\pi}} \int_{-\infty}^{-y}dt_{1}
\int_{-y}^{\infty}dt_{2}F(t_{1},t_{2}) \\
& =    \frac{1}{\sqrt{2\pi}} \int_{-\infty}^{\infty}dt_{1}  \, \int_{-\infty}^{-y}dt_{2} \,
F(t_{1},t_{2}) -\frac{1}{2\sqrt{2\pi}} \int_{-\infty}^{-y}dt_{1} \int_{-\infty}^{-y}dt_{2}F(t_{1},t_{2})\\
& =   \frac{1}{\sqrt{2\pi}} \int_{-\infty}^{\infty}dt_{1} \int_{-\infty}^{-y} \, dt_{2} \, F(t_{1},t_{2})
+ A_{4}(-y),
\end{align*}
and thus
\begin{align*}
A_{a,4}(y) &=   \frac{1}{2\sqrt{2\pi}} \int_{-\infty}^{\infty}dt_{1} \int_{-\infty}^{-y}dt_{2 } \,F(t_{1},t_{2}).
\end{align*}
From this, we apply the LHS to get
\begin{align*}
e^{- 2y^{2}}\partial_{y}\left(e^{2y^{2}} \partial_{y}A_{a,4}(y)\right)& = -\frac{\sqrt{2}}{ \sqrt{\pi}} \int_{-\8}^{\8} dx   \, e^{-2(x-y)^{2} } \Big({\rm erf} (x) - {\rm erf} \big(\sqrt{2}y \big) \Big) = {\rm erf}(\sqrt{2}y)-{\rm erf}(\sqrt{2}y/\sqrt{3}).
\end{align*}
Combining this with (\ref{E5}) proves the lemma. 


\begin{thebibliography}{10}

\bibitem{AJ81}
A. Alastuey and B. Jancovici, \emph{On the two-dimensional one-component {Coulomb} plasma},
J. Physique \textbf{42} (1981), 1--12.

\bibitem{AHM08}
Y.~Ameur, H.~Hedenmalm, and N.~Makarov, \emph{Fluctuations of eigenvalues of
  random matrices}, Duke Math. J. \textbf{159} (2011), 31--81.

\bibitem{BH08a}
B.A. Bernevig and F.D.M. Haldane, \emph{Model fractional quantum {H}all states
  and {J}ack polynomials}, Phys. Rev. Lett. \textbf{100} (2008), 246802.

\bibitem{BR09}
B.A. Bernevig and N.~Regnault, \emph{The anatomy of {A}belian and non-{A}belian
  fractional quantum {H}all states}, Phys. Rev. Lett. \textbf{103} (2009),
  206801.
  
  \bibitem{CFTW13} T.~Can, P.J.~Forrester, G.~T\'ellez and P.~Wiegmann,
  \emph{Singular behaviour at the edge of {L}aughlin states}, arXiv:1307.3334

\bibitem{CFS83}
Ph. Choquard, P.J. Forrester, and E.R. Smith, \emph{The two-dimensional
  one-component plasma at {$\Gamma = 2$}: the semiperiodic strip}, J. Stat.
  Phys. \textbf{33} (1983), 13--22.
  
  \bibitem{CW03}  
O.~Ciftja and C.~Wexler, \emph{Monte {C}arlo simulation method for {L}aughlin-like states
in a disk geometry},
Phys. Rev. B \textbf{67} (2003), 075304.


\bibitem{DMF96}  
N.~Datta and R.~Morf and R.~Ferrari, \emph{Edge of the {Laughlin} droplet},
Phys. Rev. B \textbf{53} (1996), 10906--10915.
  

\bibitem{Fo91}
P.J. Forrester, \emph{Finite size corrections to the free energy of {Coulomb}
  systems with a periodic boundary condition}, J. Stat. Phys. \textbf{63}
  (1991), 491--504.

\bibitem{Fo99}
\bysame, \emph{Fluctuation formula for complex random matrices}, J. Phys. A
  \textbf{32} (1999), L159--L163.
  
   \bibitem{Fo10}
\bysame, \emph{Log-gases and random matrices}, Princeton University Press (2010).
  
\bibitem{Fo11a}
\bysame, \emph{Spectral density asymptotics for {G}aussian and {L}aguerre $\beta$-ensembles in the exponentially small region},   J. Phys. A
  \textbf{45} (2012), 075206.
  
 
  
\bibitem{Fo11b}
\bysame, \emph{Large deviation eigenvalue density for the soft edge Laguerre and Jacobi $\beta$-ensembles },  J. Phys. A
  \textbf{45} (2012), 145201.


  
\bibitem{FM09}
P.J. Forrester and A. Mays,    \emph{A method to calculate correlation functions for $\beta = 1$
random matrices of odd size}, J. Stat. Phys. \textbf{134} (2009), 443--462.

\bibitem{Gi65} 
J. Ginibre, \emph{Statistical ensembles of complex, quaternion, and real matrices},
J. Math. Phys. \textbf{6} (1965), 440-440.

\bibitem{Ja81}
B.~Jancovici, \emph{Exact results for the two-dimensional one-component
  plasma}, Phys. Rev. Lett. \textbf{46} (1981), 386--388.

\bibitem{JMP94}
B.~Jancovici, G.~Manificat, and C.~Pisani, \emph{Coulomb systems seen as
  critical systems: finite-size effects in two dimensions}, J. Stat. Phys.
  \textbf{76} (1994), 307--330.

\bibitem{La83}
R.B. Laughlin, \emph{Anomalous quantum {H}all effect: an incompressible quantum
  fluid with fractionally charge excitations}, Phys. Rev. Lett. \textbf{50}
  (1983), 1395--1398.
  
  \bibitem{MH86}
  R.~Morf and B.I.~Halperin, \emph{Monte {C}arlo evaluation of trial wave functions for the
  fractional quantized {H}all effect: disk geometry}, Phys. Rev. B \textbf{33} (1986), 2221--2246.

\bibitem{RV07s}
B.~Rider and B.~Vir\'ag, \emph{The noise in the circular law and the {G}aussian
  free field}, IMRN \textbf{2007} (2007), rnm006.
  
  \bibitem{SWK04} L. \v{S}amaj, J. Wagner, and P. Kalinay, {\it Translation Symmetry Breaking in the One-Component Plasma on the Cylinder}, J. Stat. Phys \textbf{117}  (2004), 159--178.
  
 \bibitem{SM76} R.R. Sari and D. Merlini, {\it On the $\nu$-dimensional one-component classical plasma:
the thermodynamic limit revisited}, J. Stat. Phys. \textbf{76} (1976), 91--100.
   

\bibitem{Sh11}
S.~Shakirov, \emph{Exact solution for mean energy of 2d {Dyson} gas at $\beta =
  1$}, Phys. Lett. A \textbf{375} (2011), 984--989.

\bibitem{T13} G. T\'ellez, {\it Exactly solvable models in statistical
    mechanics of Coulomb systems},
  Rev. Acad. Colomb. Cienc. \textbf{37} (2013), 61--74.
  
\bibitem{TF99}
G.~T\'ellez and P.J. Forrester, \emph{Finite size study of the {2dOCP} at
  {$\Gamma=4$} and {$\Gamma=6$}}, J. Stat. Phys. \textbf{97} (1999), 489--521.
  
  \bibitem{TF12}
\bysame, \emph{Expanded Vandermonde Powers and sum rules for the two-dimensional one-component plasma}, J. Stat. Phys. \textbf{148} (2012), 824--855.

\bibitem{Wi12}
P.~Wiegmann, 
 \emph{Nonlinear hydrodynamics and fractionally quantized solitons at the fractional quantum Hall edge}, Phys. Rev. Lett. \textbf{108}, 206810.

\bibitem{ZW06} A.~Zabrodin and P.~Wiegmann, \emph{Large-$N$ expansion for the {2D}
{Dyson} gas},
J.~Phys.~A \textbf{39} (2006), 8933.

\end{thebibliography}

\providecommand{\bysame}{\leavevmode\hbox to3em{\hrulefill}\thinspace}
\providecommand{\MR}{\relax\ifhmode\unskip\space\fi MR }
\providecommand{\MRhref}[2]{%
  \href{http://www.ams.org/mathscinet-getitem?mr=#1}{#2}
}
\providecommand{\href}[2]{#2}

\end{document}